\documentclass[journal,comsoc]{IEEEtran}
\usepackage{flushend} 

\usepackage{graphicx}
\usepackage{epstopdf}
\usepackage{float}
\usepackage{algorithm,algorithmic}
\usepackage{array}
\usepackage{amsmath}
\usepackage{amssymb}
\usepackage{mdwmath}
\usepackage{CJK}
\usepackage{mdwtab}
\usepackage{eqparbox}
\usepackage{fixltx2e}
\usepackage{cases}
\usepackage{bm}
\usepackage{multirow}
\usepackage{psfrag}
\usepackage[usenames]{color}
\usepackage{mathtools}

\usepackage{fixmath}
\usepackage{mathdots}
\usepackage{latexsym}

\usepackage{url}
\usepackage{cite}
\ifCLASSOPTIONcompsoc
\usepackage[caption=false,font=normalsize,labelfont=sf,textfont=sf]{subfig}
\else
\usepackage[caption=false,font=footnotesize]{subfig}
\fi

\newtheorem{proposition}{Proposition}

\newcommand{\diag}{\mathop{\mathrm{diag}}}
\newcommand{\tr}{\mathop{\mathrm{Tr}}}

\newcommand{\tabincell}[2]{\begin{tabular}{@{}#1@{}}#2\end{tabular}}

\makeatletter
\renewcommand*{\@opargbegintheorem}[3]{\trivlist
      \item[\hskip \labelsep{\bfseries #1\ #2}] \textbf{(#3):}\ }
\makeatother

\begin{document}

\title{Reflecting Modulation}
\author{Shuaishuai~Guo,~\IEEEmembership{Member, IEEE,}
      Shuheng~Lv, \IEEEmembership{Student Member, IEEE,} ~Haixia~Zhang,~\IEEEmembership{Senior Member, IEEE,}
Jia~Ye,~\IEEEmembership{Student Member, IEEE,}~and Peng~Zhang,~\IEEEmembership{Member, IEEE}
\thanks{S. Guo, S. Lv and H. Zhang are  with Shandong Provincial Key Laboratory of Wireless Communication Technologies and School of Control Science and Engineering, Shandong University, Jinan 250061, China (email: shuaishuai\textunderscore guo@sdu.edu.cn; shuheng.lv@mail.sdu.edu.cn; haixia.zhang@sdu.edu.cn;).}
\thanks{J. Ye  is  with the Computer, Electrical and
Mathematical Science and Engineering Division,  King Abdullah University of Science and Technology (KAUST), Thuwal 23955-6900, Saudi Arabia (email: jia.ye@kaust.edu.sa).}
\thanks{P. Zhang is with the School of Computer Engineering, Weifang University, Weifang 261061, China (e-mail: sduzhangp@163.com).}

}

\maketitle

\begin{abstract}
Reconfigurable intelligent surface (RIS) has emerged as a promising technique for future wireless communication networks. How to reliably transmit information in a RIS-based communication system arouses much interest.  This paper proposes a reflecting modulation (RM) scheme for RIS-based communications, where both the reflecting patterns and transmit signals can carry information. Depending on that the transmitter and RIS jointly or independently deliver information, RM is further classified into two categories: jointly mapped RM (JRM) and separately mapped RM (SRM).   JRM and SRM are naturally superior to existing schemes, because the transmit signal vectors, reflecting patterns, and bit mapping methods of JRM and SRM are more flexibly designed.   To enhance transmission reliability, this paper proposes a discrete optimization-based joint signal mapping, shaping, and reflecting (DJMSR) design for JRM and SRM to minimize the bit error rate (BER) with a given transmit signal candidate set and a given reflecting pattern candidate set. To further improve the performance, this paper optimizes multiple reflecting patterns and their associated transmit signal sets in continuous fields for  JRM and SRM.  Numerical results show that JRM and SRM with the proposed system optimization methods considerably outperform existing schemes in BER.
\end{abstract}


\IEEEpeerreviewmaketitle

\section{Introduction}

\IEEEPARstart{R}{econfigurable} intelligent surface (RIS), also referred as intelligent reflecting surface (IRS),  has newly emerged as a promising technique for wireless communications to against unfavorable wireless environment\cite{di2019smart}. RIS consists of a massive number of passive reflecting units, which neither introduce too much additional noise  nor need signal processing circuits. Compared to relays, passive RIS can be realized with minimal hardware complexity and cost based on low-power and low-complexity electronic circuits \cite{Ntontin2019}. Recently, there has been a large body of literature investigating RIS on the channel estimation \cite{Nadeem2019,Taha2019,He2020,Mishra2019}, joint precoding and reflecting designs \cite{Ye2019, Huang2019,Fu2019, Yu2019}, information modulation techniques \cite{Basar2019,Basar2019a,Basar2019b,Nguyen2019}, performance analysis \cite{8319526, hu2017potential,hu2018capacity,jung2018performance,Zhang2019,Karasik2019},  hardware implementation and experimental work \cite{Tang2019,Tang2019a,Tang2019b,Tang2019c}, etc.
Among them, the investigation on RIS-based information transfer schemes arouses our special interest. Specifically, we are interested in how to reliably convey information at a fixed rate of $r$ bits per channel use (bpcu) in an $N_t\times N_r$ MIMO communication system assisted by an $N$-unit RIS.

\subsection{Prior Work on RIS-Based Information Transfer}
Various RIS-based information transfer schemes in literature can fulfill the transmission. According to the roles of RIS in various schemes, we classify prior work into four categories: RIS-aided communications (RIS-C), RIS-based backscatter communications (RIS-BC), RIS-based spatial modulation (RIS-SM),  passive beamforming and information transfer (PBIT).

\subsubsection{RIS-C} In RIS-C, RIS only reflects signals. RIS-C systems attracted the most research attention in literature including the investigation for spectral efficiency (SE)/signal power maximization, capacity/data rate optimization, security/reliable transmission analysis, channel estimation, etc. Huang \emph{et al} \cite{Huang2019} made a valuable contribution on maximizing energy efficiency by jointly designing the RIS phase rotating matrix and power allocation at the base station. The authors of \cite{Fu2019} and \cite{Yu2019} improved the power efficiency through optimizing the beamformer at the transmitter and the phase shift matrix at RIS. The primary and extended works on channel capacity were provided by Hu \emph{et al} \cite{8319526, hu2017potential}, who established the relationship between capacity per square meter surface area and the average transmit power. Later, Hu \emph{et al} further examined the degradations in capacity assuming RIS has hardware impairments in \cite{hu2018capacity}. Jung \emph{et al} \cite{jung2018performance} derived the asymptotic results of uplink data rate in a RIS-C system considering channel estimation errors and spatially correlated Rician fading with channel hardening effects. Zhang \emph{et al}\cite{Zhang2019} analyzed the RIS-aided multiple-input multiple-output (MIMO) capacity by using an alternative optimization approach.
The physical layer security of RIS-C systems also gained the attention of researchers. The recent papers \cite{cui2019secure,shen2019secrecy,yu2019enabling} have made contributions on improving the difference between the data rate at the legitimate receiver and the one at an eavesdropper. It is noteworthy that these investigations were based on Gaussian input assumption to provide the insights into communications performance bounds, but neglected the fact that finite constellation signals are the most common input to RIS-C systems. Against the background, Ye \emph{et al} investigated the joint reflecting and precoding designs to minimize the symbol error rate (SER) for RIS-C in \cite{Ye2019}. More recent attention has focused on the provision of channel estimation in RIS-C systems, which becomes challenging because of its massive number of passive elements without any signal processing capability.  The pilot training signals were first introduced by \cite{jung2019performance} to obtain the channel state information (CSI). They found the optimal pilot training length, which could maximize the asymptotic SE.  Nadeem \emph{et al} \cite{Nadeem2019} designed channel estimation protocol based on minimum mean squared error, while Taha \emph{et al} \cite{Taha2019} solved this problem by using the compressive sensing and deep learning methods. Three-stage mechanisms regarding channel estimation including sparse matrix factorization, ambiguity elimination, and matrix completion were proposed by \cite{He2020}. 
\subsubsection{RIS-BC} In RIS-BC, RIS plays the role of information modulator.
Backscatter communication is another promising communication paradigm that enables the backscatter devices to modulate the information over the ambient radio frequency (RF) signals without requiring active energy-emitting components. Specifically, a transmitter tag switches its antenna to non-reflecting or reflecting mode based on external energy sources in the ambient environment, such as WiFi, public radio, and cellular transmit power. The differences and similarities between RIS and backscatter leading to the occurrence of RIS-based backscatter communication (RIS-BC) systems appeared in recent works \cite{Basar2019a,Tang2019a, Tang2019b, Tang2019c} by Tang \emph{et al}. They have established several novel wireless communication systems by designing the hardware structure of the transmitter based on the concept of the programmable metasurface. Reflection coefficient controllable metasurface-based transmitter occurred in \cite{Tang2019a}, which could process phase modulation of the reflected electromagnetic (EM) wave directly. RF chain-free transmitter and space-down-conversion receiver were proposed in \cite{Tang2019b} based on the superior EM waves manipulation capability of programmable metasurfaces. In \cite{Tang2019c}, they have designed a novel transmitter without filter, wideband mixer and power amplifier on the concept of the new programmable metasurface architecture. From the perspective of design architectures and preliminary experimental results, all proposed wireless programmable metasurfaces networks were verified to achieve low hardware complexity, low cost, and high energy efficiency. Another example of RIS-BC is RIS-based space shift keying (RIS-SSK) proposed by Basar in \cite{Basar2019a}, where the transmitter only radiates a carrier signal.
\subsubsection{RIS-SM} In RIS-SM, RIS both reflects and carries the information.
During the past few years, there has been a growing interest in using the channel index for information modulation. A notable example is a spatial modulation (SM) \cite{Mesleh2008,Renzo2014,Guo2019a,Guo2019}, which could simplify the transceiver architecture and increase the energy efficiency. The intention propagation environment controllability of RIS boosted the research on RIS-SM systems \cite{Basar2019,Basar2019a,Basar2019b,Karasik2019}. The preliminary contributions in this field appeared in \cite{Basar2019}, who investigated the effect of modulation orders and blind phases on the error performance of the RIS-based communication system. The phenomenon that the RIS-based scheme experiences degradation of the error performance as the modulation order increases was found in \cite{Basar2019b}. However, it also showed that RIS could take advantage of large numbers of reflecting elements to counteract the detrimental effect of increasing the modulation order.  Considering maximum energy-based suboptimal and exhaustive search-based optimal detectors, theoretical analysis, and computer simulation results on average bit error probability were provided to validate the potential of RIS-assisted index modulation schemes on improving the SE and data rates with remarkably low error rate. \cite{Karasik2019} showed that jointly encoding RIS and the transmitter signals outperforms the RIS-C transmission. It is worth mentioning only the reflecting patterns that steer the beam to a single receive antenna were adopted for data communication in RIS-SM \cite{Basar2019a}. Using such reflecting patterns cannot benefit from the receive diversity gain.
\subsubsection{PBIT} In PBIT, RIS helps the information delivery from the transmitter to the receiver and has its own information to be transmitted.
Most recently, PBIT was firstly proposed and investigated by Yan \emph{et al} in \cite{yan2019passive}. They maximized the average received signal-to-noise ratio (SNR) assuming RIS data adopting SM. In detail, the RIS information is carried by the ON/OFF states of the reflecting elements, while passive beamforming is achieved by adjusting the phase shifts of the activated reflecting elements. The main difference between RIS-SM and PBIT is that the transmitter and RIS in RIS-SM can jointly encode the information while the transmitter and RIS in PBIT cannot because they does not share the information to be transmitted.  It should be noted that the reflecting units can only be ON or OFF to carry information in PBIT, which limits the feasible reflecting patterns.

To conclude, the literature identifies that RIS-based information transfer can rely on the radiated signals at the transmitter, the reflecting patterns at the RIS  or both for data delivery. The transmitter and RIS can either independently or jointly deliver information. The adopted reflecting patterns can be the patterns steering the beam to a single receive antenna or just the ON/OFF states of each RIS unit. The dimension of the radiated signals can be one-dimensional and multi-dimensional. All these schemes call a unified transmission model for a fair comparison. Besides, there has been no detailed investigation of system optimization involving signal mapping, shaping and reflecting pattern design. All these motivated our work. It is hoped that our work will contribute to a deeper understanding of RIS-based information transfer.

\subsection{Our Work and Contributions}
\begin{itemize}
\item In this paper, we propose a reflecting modulation (RM) scheme for RIS-based communications. Depending on that the transmitter and RIS jointly or independently deliver information, RM is further classified into two categories: jointly mapped RM (JRM) and separately mapped RM (SRM).  JRM and SRM (JRM\&SRM) can cover the existing schemes, i.e., RIS-C, RIS-BC, RIS-SM and PBIT. Since the transmit signal vectors, reflecting patterns, and  bit mapping methods in JRM\&SRM are more flexibly designed, JRM\&SRM are naturally superior to existing schemes.  BER analysis of JRM\&SRM is included, which will generate fresh insight into how the signal mapping, shaping, and reflecting affect the system BER.
\item To enhance transmission reliability, this paper proposes a discrete optimization-based joint signal mapping, shaping, and reflecting (DJMSR) design for  JRM\&SRM to minimize the system BER with  a given transmit signal candidate set and a given reflecting pattern candidate set. We compare DJMSR with an exhaustive search method  in complexity and performance to validate its effectiveness. 
 \item To further improve the performance, this paper proposes a continuous optimization-based joint signal mapping, shaping, and reflecting (CJMSR) design for JRM\&SRM. Using an alternative optimization approach, we iteratively optimize the signal shaping and reflecting in continuous fields. In the reflecting design with given transmit signal sets, multiple reflecting patterns for reflecting and carrying information are jointly optimized.  In the signal shaping design with given reflecting patterns, the transmit signal sets for all reflecting patterns are jointly optimized. 
 \item Comprehensive numerical results are presented. The proposed JRM\&SRM are compared to RIC-C, RIS-BC, RIS-SM, and PBIT to validate their superiority. The effectiveness of DJMSR\&CJMSR is validated in various system setups. The impact of channel estimation errors on the performance of the JRM\&SRM with DJMSR and CJMSR (DJMSR\&CJMSR) is studied. Moreover, implementation challenges and future directions regarding JRM\&SRM are also discussed.

\end{itemize}

\begin{table*}[t]
\centering
\caption{Separately Mapped RM (SRM)}\label{tab1}
\begin{tabular}{ | c | c |c|c|c|}
    \hline
     \tabincell{c}{Bits at the\\  transmitter}
  & \tabincell{c}{Bits at \\  the RIS}&\tabincell{c}{Activated \\pattern}&\tabincell{c}{Transmitted \\signal}&\tabincell{c}{Transmitted \\signal sets}\\
\hline
00 & 0&$\mathbf{\Phi}_1$&$\textbf{x}_1$&\multirow{4}*{$\mathcal{X}_1=\{\textbf{x}_1,\textbf{x}_2,\textbf{x}_3,\textbf{x}_4\}$}\\
\cline{1-4}
01 & 0&$\mathbf{\Phi}_1$&$\textbf{x}_2$&\\
\cline{1-4}
10 & 0&$\mathbf{\Phi}_1$&$\textbf{x}_3$&\\
\cline{1-4}
11 & 0&$\mathbf{\Phi}_1$&$\textbf{x}_4$&\\
\cline{1-5}
00 & 1&$\mathbf{\Phi}_2$&$\textbf{x}_1$&\multirow{4}*{$\mathcal{X}_2=\{\textbf{x}_1,\textbf{x}_2,\textbf{x}_3,\textbf{x}_4\}$}\\
\cline{1-4}
01 & 1&$\mathbf{\Phi}_2$&$\textbf{x}_2$&\\
\cline{1-4}
10 & 1&$\mathbf{\Phi}_2$&$\textbf{x}_3$&\\
\cline{1-4}
11 & 1&$\mathbf{\Phi}_2$&$\textbf{x}_4$&\\
\hline
    \end{tabular}
\end{table*}

\begin{table*}[t]
\centering
\caption{Jointly Mapped RM (JRM)}\label{tab1}
\begin{tabular}{ | c  |c|c|c|}
    \hline
Bits &Activated pattern&Transmitted signal&Transmitted signal sets\\
\hline
000&$\mathbf{\Phi}_1$&$\textbf{x}_1$&\multirow{3}*{$\mathcal{X}_1=\{\textbf{x}_1,\textbf{x}_2,\textbf{x}_3\}$}\\
\cline{1-3}
001&$\mathbf{\Phi}_1$&$\textbf{x}_2$&\\
\cline{1-3}
010&$\mathbf{\Phi}_1$&$\textbf{x}_3$&\\
\cline{1-4}
011&$\mathbf{\Phi}_2$&$\textbf{x}_2$&\multirow{3}*{$\mathcal{X}_2=\{\textbf{x}_2,\textbf{x}_3,\textbf{x}_4\}$}\\
\cline{1-3}
100&$\mathbf{\Phi}_2$&$\textbf{x}_3$&\\
\cline{1-3}
101&$\mathbf{\Phi}_2$&$\textbf{x}_4$&\\
\hline
110&$\mathbf{\Phi}_3$&$\textbf{x}_1$&\multirow{2}*{$\mathcal{X}_3=\{\textbf{x}_1,\textbf{x}_4\}$}\\
\cline{1-3}
111&$\mathbf{\Phi}_3$&$\textbf{x}_4$&\\
\hline
    \end{tabular}
\end{table*}

\subsection{Paper Organization}
The rest of this paper proceeds as follows. Section II depicts the system model, including the mapping types, the signal model and the BER analysis. Section III introduces the system optimization involving signal mapping, shaping and reflecting in use of a given transmit signal candidate set and a given reflecting pattern candidate set. Section IV extends the optimization in continuous fields. Section V presents the numerical results. Section VI points out the implementation challenges and future research directions.  
Conclusions are drawn in the last section. 
\subsection{Notations}
Throughout this paper, the term $x$ refers to a scalar; $\textbf{x}$ represents a vector; and $\textbf{X}$ denotes a matrix. The terms ${\left\| \textbf{x} \right\|_0}$, ${\left\| \textbf{x} \right\|_2}$, ${\left\| \textbf{x} \right\|_p}$, ${\left\| \textbf{x} \right\|_{\infty}}$ will be used in this paper to refer to $\ell_0$, $\ell_2$, $\ell_p$ and $\ell_{\infty}$ norms. $(\textbf{X})_{m,n}$ stands for the entry of $\textbf{X}$ located at $m$-th row and $n$-th column. 
$\diag{(\textbf{x})}$ means a diagonal matrix with diagonal elements being $\textbf{x}$. $\diag\{\textbf{X}\}$ represents a vector drawn from the diagonal elements from $\textbf{X}$. $(\cdot)^T$, $(\cdot)^C$, $(\cdot)^H$ refer to the transpose, conjugate and conjugate-transpose operators, respectively. $\odot$ and $\otimes$ stands for the Hadamard and Kronecker products, respectively. $\textbf{1}_{m\times n}$ represents a all-one matrix with $m$ rows and $n$ columns. $\ln$ represents the Natural logarithm.
$\mathbb{C}$ and $\mathbb{R}$ stand for the complex and the real domains. $\textbf{I}_N$ denotes an $N\times N$ identity matrix. $\left\lfloor \cdot \right\rfloor $ denotes the floor operation. $Q(\cdot)$ stands for the tail distribution function of the standard normal distribution. $\mathcal{X}$
is a set and $\left|\mathcal{X}\right|$ represents the size of set $\mathcal{X}$. $\emptyset$ refers to an empty set.  $\left(n\atop m\right)$ is a binomial coefficient.  $\Re(x)$ and $\Im(x)$ denote the functions to take the real and imaginary part of $x$.
\section{System Model}
\begin{figure}[t]
  \centering
  \includegraphics[width=0.4\textwidth]{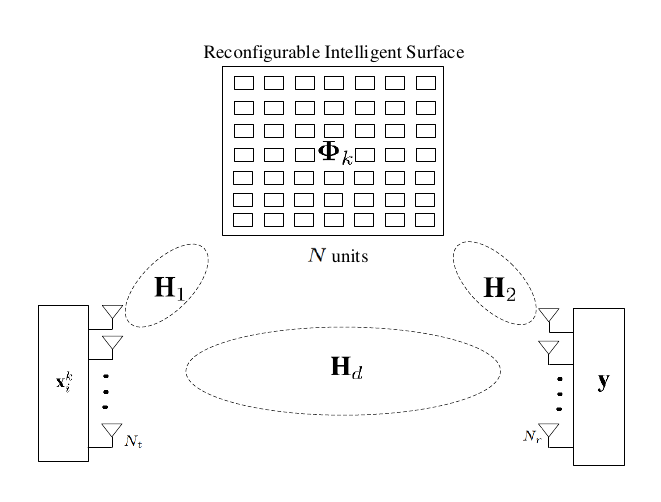}\\
 \caption{A RIS-assisted ($N_t$, $N_r$, $N$, $r$) MIMO communication system.}
  \label{System_Model}
\end{figure}
In this paper, we consider a RIS-based ($N_t$, $N_r$, $N$, $r$) MIMO communication system as illustrated in Fig. \ref{System_Model}, where $N_t$, $N_r$ represent the numbers of transmit and receive antennas at the transceivers;  $N$ stands for the number of RIS units; and $r$ is the target transmit rate in bit per channel use (bpcu). Let $L=2^r$ and there are $L$ possible bit sequences of length $r$ in a channel use. In the proposed RM scheme, similar to \cite{Basar2019,Basar2019a,Basar2019b,Nguyen2019},  a sequence $\textbf{b}_l$ of $r$ bits is conveyed per channel use not only by the index of the transmit signal vector $\textbf{x}_i^k\in \mathbb{C}^{N_t\times 1}$ but also by the index of the reflecting patterns $\mathbf{\Phi}_k\in\mathbb{C}^{N\times N}$, i.e., the tuple $(i,k)$. In the following, $\mathcal{X}_k=\{\textbf{x}_i^k\}$ denotes the transmit signal set when $\mathbf{\Phi}_k$ is activated; $\mathcal{X}$ denotes the transmit signal candidate set with size $|\mathcal{X}|=M$; and $\Psi$ represents the reflection pattern candidate set with size $|\Psi|=K$. Based on above denotations, we have $\textbf{x}_i^k\in\mathcal{X}_k\subseteq\mathcal{X}$ and $\mathbf{\Phi}_k\in\Psi$. Moreover, due to the specific nature of RIS units, $\mathbf{\Phi}_k\in\Psi$ is a diagonal matrix with each diagonal entry having modulus $1$ or $0$, where $0$ represents the OFF state. Mathematically, $(\mathbf{\Phi}_k)_{n,n}\in\{0\}\cup\{e^{j\theta},\theta\in\mathbb{R}\}$. Since $\textbf{x}_i^k$ is a general multi-dimensional signal, $\mathbf{\Phi}_k$ is a general reflecting pattern, and the adopted transmit signal sets corresponding to $\{\mathbf{\Phi}_k\}$ can be the same or not in the proposed transmission model, it is obvious that  the existing RIS-SM and PBIT \cite{Basar2019,Basar2019a,Basar2019b,Nguyen2019} can be regarded as special realizations of the proposed RM.  Moreover, RIS-C and RIS-BC can also be regarded as special realizations of RM by setting $|\Psi|=1$ and $|\mathcal{X}|=1$, respectively.

\subsection{Mapping Types}
The proposed RM is suitable for two application scenarios. In the first scenario, the RIS and the transmitter do not share the information bits and they have to independently transmit their information similarly to the PBIT in \cite{yan2019passive}.  In this scenario, the total $r$ bits can only be separately mapped to $i$ and $k$. Thus, we refer to such a RIS-based information transfer as SRM. In SRM, the number of information bits mapped to $i$ and $k$ are denoted by $r_1$ and $r_2$, respectively. Since the number of bits has to be integer, consequently the numbers of adopted transmit signals and reflecting patterns are limited to be a power of two\footnote{It should be mentioned that even though there are some methods such as fractional bit mapping method \cite{Serafimovski2010} that can release the number constraint, such methods typically suffered from severe error propagation and high detection complexity.}. In our work, we use $M_c=2^{r_1}$ and $K_c=2^{r_2}$ to represent the numbers of transmit signals and reflecting patterns that can be activated. And we have $L=M_cK_c$. Moreover, since the transmitter does not know the information to be transmitted at the RIS, thus the employed transmit signal sets for different reflecting pattern activations are the same. SRM is much similar to the concept of PBIT, but with more flexible reflecting patterns.

In the second scenario, the RIS and the transmitter jointly transmit the information bits. In other words, the information bits can be jointly mapped to the tuple $(i,k)$, similarly to the jointly mapped SM in \cite{Guo2016a, Guo2017}. In this paper, we refer to such a RIS-based information transfer as JRM. To show the differences between SRM and JRM, an example is provided, where the candidate set for reflecting patterns $\Psi=\{\mathbf{\Phi}_1,\mathbf{\Phi}_2,\mathbf{\Phi}_3\}$ and the candidate set for the transmitted signals $\mathcal{X}=\{\textbf{x}_1,\textbf{x}_2,\textbf{x}_3,\textbf{x}_4\}$. The bit mapping methods of SRM and JRM are given in Tables I and II, respectively.
From the tables, it can be observed that SRM  can only use a power of two reflecting patterns, while JRM can use a more flexible $(i,k)$ tuple set with size $L$. In JRM, the transmit signal sets for each reflecting  pattern activation can be the same or not. To show the differences, we denote $\textbf{x}_i^k$ as the $i$-th transmit signal when activating $\mathbf{\Phi}_k$. In SRM, $\{\textbf{x}_i^k\}$ are the same for any $\mathbf{\Phi}_k$ that can be activated. In the given example,  $\mathcal{X}_1=\mathcal{X}_2=\mathcal{X}$ and $\mathcal{X}_3=\emptyset$ in SRM. From this point, SRM can be regarded as a special case of JRM.   Therefore, if the transmitter and the RIS do not independently deliver the information, JRM is preferred, since the optimized JRM is surely superior to the optimized SRM. Also, we remark that given the same $\mathcal{X}$ and $\Psi$, the optimized JRM is naturally superior to RIS-C, RIS-BC, RIS-SM and PBIT. This is because RIS-C, RIS-BC, RIS-SM and PBIT are special realizations of JRM and the globally optimized solution will surely achieve better performance than the optimized solutions with additional constraints.

\subsection{Signal Model}
Let $l$ be the index of the activated tuple $(i,k)$ corresponding to the bit sequence $\textbf{b}_l$.  For either SRM or JRM, when $\mathbf{\Phi}_k$ and $\textbf{x}_i^k$ are activated to deliver $\textbf{b}_l$,
the receiver signal vector $\textbf{y}\in \mathbb{C}^{N_r\times 1}$ can be expressed by
\begin{equation}
\textbf{y}=(\textbf{H}_d+\textbf{H}_2\mathbf{\Phi}_k\textbf{H}_1)\textbf{x}_i^k+\textbf{n},
\end{equation}
as shown in Fig. \ref{System_Model}, where  $\textbf{H}_d\in \mathbb{C}^{N_r\times N_t}$ denotes channel matrix of the direct link; 
$\textbf{H}_2\in\mathbb{C}^{N_r\times N}$ represents the channel matrix  between the RIS and the receive antennas;  $\textbf{H}_1\in\mathbb{C}^{N\times N_t}$ represents the channel matrix between the transmit  and the RIS; and $\textbf{n}\in \mathbb{C}^{N_r\times 1}$ stands for the additive complex Gaussian noise with zero mean and variance $\sigma^2\textbf{I}_{N_r}$. In the paper, it is assumed that all channels are perfectly known by the transceivers as well as the RIS.

With $\textbf{H}_d$, $\textbf{H}_2$ and $\textbf{H}_1$ being known by the receiver,  maximum likelihood (ML) detection can be performed by
\begin{equation}
\begin{split}
(\hat{i},\hat{k})&=\arg\max_{\textrm{all legitimate}(i,k)}p_{\mathbf{Y}}(\mathbf{y}|\textbf{x}_i^{k},\mathbf{\Phi}_k,\textbf{H}_d,\textbf{H}_2,\textbf{H}_1)\\
&=\arg\min_{\textrm{all legitimate}(i,k)} ||\textbf{y}-(\textbf{H}_d+\textbf{H}_2\mathbf{\Phi}_k\textbf{H}_1)\textbf{x}_i^k||^2_2,
\end{split}
\end{equation}
which is deduced from
\begin{equation}
\begin{split}
p_{\mathbf{Y}}(\mathbf{y}|\textbf{x}_i^k,\mathbf{\Phi}_k,&\textbf{H}_d,\textbf{H}_2,\textbf{H}_1)\propto\exp(-||\textbf{y}-(\textbf{H}_d+\textbf{H}_2\mathbf{\Phi}_k\textbf{H}_1)\textbf{x}_i^k||^2_2).
\end{split}
\end{equation}
\subsection{BER Analysis}
According to \cite{Tse2005}, the BER of JRM\&SRM can be analyzed to be upper bounded by
\begin{equation}\label{BER}
P_e\leq \overline{P_e}=\frac{1}{Lr}\sum_{l=1}^{L}\sum_{\hat{l}=1,\hat{l}\neq l}^{L}D_{\rm{HD}}(\textbf{b}_l,\textbf{b}_{\hat{l}})P_{\rm{PEP}}(l\rightarrow\hat{l}),
\end{equation}
where $D_{\rm{HD}}(\textbf{b}_l,\textbf{b}_{\hat{l}})$ represents the Hamming distances between $\textbf{b}_l$ and $\textbf{b}_{\hat{l}}$; $P_{\rm{PEP}}(l\rightarrow\hat{l})$ represents the pairwise error probability. In (\ref{BER}), $P_{\rm{PEP}}(l\rightarrow\hat{l})$ can be calculated by
\begin{equation}
\begin{split}
P(l\rightarrow\hat{l})=Q\left(\sqrt{\frac{D_{\rm{ED}}(l,\hat{l})^2}{2\sigma^2}}\right),
\end{split}
\end{equation}
where
\begin{equation}
D_{\rm{ED}}(l,\hat{l})^2=||(\textbf{H}_d+\textbf{H}_2\mathbf{\Phi}_k\textbf{H}_1)\textbf{x}_i^k-(\textbf{H}_d+\textbf{H}_2\mathbf{\Phi}_{\hat{k}}\textbf{H}_1)\textbf{x}_{\hat{i}}^{\hat{k}}||_2^2,
\end{equation}
and $D_{\rm{ED}}(l,\hat{l})$ represents the Euclidean distances between the two noise-free received signal vectors.

\section{Discrete Optimization-Based Joint Signal Mapping, Shaping and Reflecting Design (DJMSR)}

In this paper, one key concern is how to design the system to minimize the system BER. In the section, we will present the joint signal mapping, shaping, reflecting  design  to minimize the BER upper bound $\overline{P_e}$ with a given transmit signal candidate set $\mathcal{X}$ and a given reflecting pattern candidate set $\Psi$. 
\subsection{Problem Formulation}
For JRM, we refer to the bijective mapping rule of sequences $\{\textbf{b}_l\}$ to tuples $\{(i,k)\}$ as $\Gamma$.
Observing the expression of  $\overline{P_e}$ in (\ref{BER}), we find that the BER upper bound is jointly determined by the Hamming distances $\{D_{\rm{HD}}(\textbf{b}_l,\textbf{b}_{\hat{l}})\}$ and the Euclidean distances $\{D_{\rm{ED}}(l,\hat{l})\}$. Among them, the Hamming distances $\{D_{\rm{HD}}(\textbf{b}_l,\textbf{b}_{\hat{l}})\}$ are only affected by the mapping rule, i.e., $\Gamma$, while the Euclidean distances $\{D_{\rm{ED}}(l,\hat{l})\}$ are jointly affected by the adopted reflecting patterns $\mathbf{\Phi}_1,\mathbf{\Phi}_2,\cdots,\mathbf{\Phi}_K$ and the corresponding transmit signal sets $\mathcal{X}_1,\mathcal{X}_2,\cdots, \mathcal{X}_K$. Based on these denotations, given a transmit signal candidate set $\mathcal{X}$ and a reflecting pattern candidate set $\Psi$, the DJMSR optimization problem for JRM can be formulated to be 
\begin{subequations}
\begin{align}
(\textbf{OP1}):~\mathrm{Given}: &~\textbf{H}_d,\textbf{H}_1,\textbf{H}_2,\sigma,\mathcal{X},\Psi,L\notag\\
\mathrm{Find}:&~\Gamma,~\mathcal{X}_1,\mathcal{X}_2,\cdots, \mathcal{X}_K\notag\\
\mathrm{Minimize}:&~\overline{P_e}\notag\\
\mathrm{Subject~to}:&~\Gamma \textrm{~is bijective,}\label{P1a}\\
&~\mathcal{X}_k\subseteq\mathcal{X},~k=1,2,\cdots,K,\label{P1b}\\
&~|\mathcal{X}_k|\geq 0,k=1,2,\cdots,K,\label{P1c}\\
&~\sum_{k=1}^{K}|\mathcal{X}_k|=L,\label{P1d}\\
&~\frac{1}{L}\sum_{k=1}^{K}\sum_{\textbf{x}_i^k\in\mathcal{X}_k}||\textbf{x}_i^k||^2_2=1\label{P1e},
\end{align}
\end{subequations}
where (\ref{P1a}) is the constraint for the mapping rule; (\ref{P1b}) limits all the transmitted signal sets corresponding to different reflecting patterns to be subsets of $\mathcal{X}$;  (\ref{P1c}) and (\ref{P1d}) represents the set size constraints for the transmit signal sets; (\ref{P1e}) stands for the normalized power constraint for all transmit signals. It is  noteworthy that even though the optimization variables does not include $\{\mathbf{\Phi}_k\}$, the reflecting patterns can be optimized, because $|\mathcal{X}_k|$ being nonzero or not will determine $\mathbf{\Phi}_k$ being used or not.

For SRM, we denote the bits-to-transmit-signal mapping rule as $\Gamma_1$ and the bits-to-reflecting-pattern mapping rule as $\Gamma_2$. Since the adopted transmit signal sets for each reflecting pattern activation have to be the same,  the number of feasible solutions of 
$\mathcal{X}_1,~\mathcal{X}_2,\cdots, \mathcal{X}_K$ is greatly reduced. Recall that $M_c$, $K_c$ are used to denote the numbers of the transmit signals and reflecting patterns that can be activated.  It means that there are $K_c$ nonzeros in a vector $\textbf{n}=[|\mathcal{X}_1|,|\mathcal{X}_2|,\cdots,|\mathcal{X}_K|]^T$, i.e., $||\textbf{m}||_0=K_c$ and let $n_1,n_2,\cdots,n_{K_c}$ be the nonzero positions in $\textbf{n}$.
Based on these denotations, the DJMSR optimization in SRM can be formulated to be\begin{subequations}
\begin{align}
(\textbf{OP2}):~\mathrm{Given}: &~\textbf{H}_d,\textbf{H}_1,\textbf{H}_2,\sigma,\mathcal{X},\Psi, K_c, r_1\notag\\
\mathrm{Find}:&~\Gamma_1,\Gamma_2,\mathcal{X}_1,\mathcal{X}_2,\cdots, \mathcal{X}_K\notag\\
\mathrm{Minimize}:&~\overline{P_e}\notag\\
\mathrm{Subject~to}:&~\Gamma_1, \Gamma_2 \textrm{~are bijective,}\label{P2a}\\
&~||\textbf{n}||_0=K_c,\label{P2b}\\
&~\mathcal{X}_{n_1}=\mathcal{X}_{n_2}=\cdots=\mathcal{X}_{n_{K_c}}\subseteq\mathcal{X},\label{P2c}\\
&~|\mathcal{X}_{n_1}|=M_c,\label{P2d}\\
&~\frac{1}{M_c}\sum_{\textbf{x}_i\in\mathcal{X}_{n_1}}||\textbf{x}_i||_2^2=1,\label{P2e}
\end{align}
\end{subequations}
where (\ref{P2a}) is the constraint for the mapping rules; (\ref{P2b}) limits the number of the reflecting patterns that can be activated; (\ref{P2c}) limits the transmit signal sets chosen from the candidate set for different reflecting patterns are the same; (\ref{P2d}) is the size constraint for the transmit signal set; (\ref{P2e}) is the normalized power constraint for all transmit signals.

\subsection{Design Procedure}
In this subsection, we will analyze the formulated problems (\textbf{OP1}) and (\textbf{OP2}) to optimize the signal mapping, shaping, and reflecting for JRM\&SRM. 
According to [35, \emph{Lemma 1}], it is known that the BER is more sensitive to the Euclidean distances  $\{D_{\rm{ED}}(l,\hat{l})\}$ than the Hamming distances $\{D_{\rm{HD}}(\textbf{b}_l,\textbf{b}_{\hat{l}})\}$ especially in the high SNR regime. Based on the fact, one can firstly optimize the signal shaping and reflecting that affects $\{D_{\rm{ED}}(l,\hat{l})\}$ with omitting the impact of signal mapping by setting  $D_{\rm{HD}}(\textbf{b}_l,\textbf{b}_{\hat{l}})=1,~\forall l\neq \hat{l},$ and then optimize the signal mapping that affects $\{D_{\rm{HD}}(\textbf{b}_l,\textbf{b}_{\hat{l}})\}$.

  In JRM, the joint signal shaping and reflecting optimization  is indeed a subset selection problem. The equivalence can be explained by an understanding of JRM that there are $KM$ feasible $(i,k)$ tuples and $L$ out of them are chosen for data transmission. Thus, one can perform exhaustive search with complexity $\mathcal{O}
\left(\left(KM\atop L\right)\mathcal{C}_{\overline{P_e}}\right)$, where $\mathcal{C}_{\overline{P_e}}$ represents the computational complexity required for computing the objective function $\overline{P_e}$. The computational complexity is sometimes prohibitive, for instance, when $K=10$, $M=10$, $L=64$, $\left(KM\atop L\right)=\left(100\atop L\right) \approx 2.0\times 10^{27}$. To reduce the complexity, we propose to use a stepwise depletion procedure, which can be described as follows. The stepwise depletion is started with initializing the legitimate tuple set to have all feasible tuples. Pick up a tuple and compute the $\overline{P_e}$ of the  rest tuple set with the average transmit power being normalized. Then drop one of the tuples whose rest tuple set has the minimum $\overline{P}_e$. Then its rest tuple with the minimum $\overline{P}_e$ become the legitimate tuple set. In each iteration, the number of tuples in the legitimate tuple set is reduced by $1$. Repeat the process step by step until the legitimate tuple set size being $L$.
 
 In SRM, the joint signal shaping and reflecting optimization is to solve two subset selection problems, which are selecting $K_c=2^{r_2}$ reflecting patterns that can be activated from $\Psi$ and selecting $M_c=2^{r_1}$ legitimate transmit signals from $\mathcal{X}$. The exhaustive search will induce the computational complexity of $\mathcal{O}\left(\left(K\atop K_c\right)\left(M\atop M_c\right)\mathcal{C}_{\overline{P_e}}\right)$. To facilitate the low-complexity implementation, we propose a low-complexity two-step optimization approach. In the first step, we select $M_c$ legitimate signal vectors in $\mathcal{X}$. In the selection, we propose to use a zero-embedding stepwise depletion procedure. In detail, we first embed a zero vector $\mathbf{0}\in\mathbb{C}^{N_t\times 1}$ in $\mathcal{X}$ to generate $\hat{\mathcal{X}}_c=\{\textbf{0}\}\cup\mathcal{X}$ and then perform a stepwise depletion procedure to repeatedly discard the signal vector expect $\textbf{0}$ that leads to unfavorable mutual Euclidean distances until the legitimate signal set size being $M_c+1$. Then by removing $\textbf{0}$ from the set, we can obtain the final legitimate transmit signal set of size $M_c$, denoted by $\mathcal{X}_c$. The zero embedding approach is to avoid selecting the signal vector that is close to the all-zero vector $\textbf{0}$. Even though a vector close to $\textbf{0}$ in $\mathcal{X}_c$ may not affect the mutual Euclidean distances in $\mathcal{X}_c$, it will severely damage the detection of the activated reflecting patterns. This selection of legitimate transmit signal set can be conducted off-line, since the selection does not involve the CSI. In the second step, we select the $K_c$ legitimate reflecting patterns from $\Psi$. An original stepwise depletion procedure can be adopted. That is, repeatedly discarding one of the reflecting patterns that lead to unfavorable $\overline{P_e}$ until the  set size being $K_c$.

 After the signal set and the reflecting pattern set are settled, Pseudo Gray mapping  \cite{zeger1990pseudo} can be adopted for JRM\&SRM to further reduce the BER.  Pseudo Gray mapping is a general mapping method for any signal constellation, which can be obtained by a low-complexity binary switching algorithm (BSA) \cite{Guo2016,Guo2017a}. BSA starts with an initial mapping. Then exchanging the bit labels of any two tuples generates a new mapping. Comparing them in terms of $\overline{P_e}$, we choose the one of two mappings with better performance. Repeating the process until all the labels of all tuples are exchanged, then the Pseudo-Gray mapping is obtained. To summary, we list the DJMSR for JRM and SRM in Algorithms 1 and 2.
\begin{algorithm}[t]
\caption{DJMSR Design for JRM}
\label{10}
\begin{algorithmic}[1]
\STATE {\textbf{Input:} $\textbf{H}_d,\textbf{H}_1,\textbf{H}_2,\sigma,\mathcal{X},\Psi,L$. }
\STATE{Select the legitimate tuple set by performing the stepwise depletion procedure, which is equivalent to optimizing $\mathcal{X}_1,\mathcal{X}_2,\cdots, \mathcal{X}_K$ directly.}
\STATE{Perform BSA \cite{Guo2016} to obtain Pseudo Gray mapping for $\Gamma$.}
\STATE{Output $\Gamma$ and $\mathcal{X}_1,\mathcal{X}_2,\cdots, \mathcal{X}_K$.}
\end{algorithmic}
\end{algorithm}
\begin{algorithm}[t]
\caption{DJMSR Design for SRM}
\label{10}
\begin{algorithmic}[1]
\STATE {\textbf{Input:} $\textbf{H}_d,\textbf{H}_1,\textbf{H}_2,\sigma,\mathcal{X},\Psi, K_c, M_c$. }
\STATE{Select the legitimate transmit signal set by performing a zero-embedding stepwise depletion procedure.}
\STATE{Select the legitimate reflecting pattern set by performing a stepwise depletion procedure.}
\STATE{Perform BSA \cite{Guo2016} to generate Pseudo Gray mapping for $\Gamma_1$ and $\Gamma_2$.}
\STATE{Output $\Gamma_1$, $\Gamma_2$, and $\mathcal{X}_1,\mathcal{X}_2,\cdots, \mathcal{X}_K$.}
\end{algorithmic}
\end{algorithm}

 \subsection{Computational Complexity and Performance Analysis}
In the DJMSR for JRM, the computational complexity of the joint shaping and reflecting as well as the signal mapping in use of BSA \cite{Guo2016} are all dominated by the computation of the objective function $\overline{P_e}$. 
In the stepwise depletion procedure for joint shaping and reflecting, finding the tuple to be dropped needs to compute the $\overline{P_e}$ regarding $t$ tuples by $t$ times, where $t$ denotes of the number of tuples in the legitimate tuple set.
Computing $\overline{P_e}$ once needs to compute the $t$ noise-free received signal vectors and their mutual Euclidean distances,  which requires $\mathcal{O}\left[tN^2(N_r+N_t)+t^2N_r\right]$ multiplications. Multiplying it by $t$ times yields $\mathcal{O}\left(\sum_{t=L+1}^{KM}t^2N^2(N_r+N_t)+t^3N_r\right)$. Adding the complexity of BSA \cite{Guo2016}, which requires around $L^2$ times of $\overline{P_e}$ with $L$ tuples that needs $\mathcal{O}\left[L^3N^2(N_r+N_t)+L^4N_r\right]$. Jointly considering all above and the relation $KM\ge L$, the complexity order of DJMSR for JRM is 
\begin{equation}
\mathcal{C}_{\textrm{DJMSR}}^{\textrm{JRM}}=\mathcal{O}[K^3M^3N^2(N_r+N_t)+K^4M^4N_r].
\end{equation}

Using exhaustive search (ES)-based signal shaping and reflecting for JRM with Pseudo Gray mapping induces a complexity of $\mathcal{O}\left(\left(KM\atop L\right)\mathcal{C}_{\overline{P_e}}\right)$. Based on above analysis, the aggregate computational complexity  of the ES method in the number of multiplications can be written as
\begin{equation}
\begin{split}
&\mathcal{C}_{\textrm{ES}}^{\textrm{JRM}}=\mathcal{O}\left[\left(\left(KM\atop L\right)+L^2\right)\left(LN^2(N_r+N_t)+L^2N_r\right)\right].
\end{split}
\end{equation}
Similarly, we can analyze the computation orders of DJMSR and ES  for SRM to be
\begin{equation}
\mathcal{C}_{\textrm{DJMSR}}^{\textrm{SRM}}=\mathcal{O}[K^3M_cN^2(N_r+N_t)+K^4M_c^2N_r],
\end{equation}
and 
\begin{equation}
\begin{split}
&\mathcal{C}_{\textrm{ES}}^{\textrm{SRM}}=\mathcal{O}\left[\left(\left(K\atop K_c\right)\left(M\atop M_c\right)+L^2\right)\left(LN^2(N_r+N_t)+L^2N_r\right)\right].
\end{split}
\end{equation}

To quantify the performance improvement brought by the proposed DJMSR compared to a benchmark system with given signal vector sets, reflecting pattern set and with a natural mapping scheme, the SNR gain for achieving a same BER in the high SNR regime can be computed as \cite{Guo2016} 
\begin{equation} \label{eq13}
\gamma_G=-\lim_{\textrm{SNR}\rightarrow \infty}\frac{10\log_{10}\frac{P_e^{\textrm{D}}}{P_e^{\textrm{B}}}}{N_r} ~(\textrm{dB}),
\end{equation}
where $P_e^{\textrm{D}}$ represents the BER of the system with the proposed DJMSR and $P_e^{\textrm{B}}$ represents the BER of the benchmark system.
In the high SNR, the BER can be approximated to be
\begin{equation}\label{eq14}
P_e\approx\lambda \frac{\overline{D}_{\rm{HD}}^{\min}}{r}Q\left(\frac{D_{\rm{ED}}^{\min}(l,\hat{l})}{\sqrt{2}\sigma}\right)\approx \lambda \frac{\overline{D}_{\rm{HD}}^{\min}}{2r}\exp\left(-\frac{{D_{\rm{ED}}^{\min}(l,\hat{l})}^2}{4\sigma^2}\right),
\end{equation}
where $\lambda$ represents the number of closest pairs; $\overline{D}_{\rm{HD}}^{\min}$ stands for the average Hamming distance between the bit sequences mapped to the closest pairs; and $D^{\min}_{\rm{ED}}$ refers to the Euclidean distance of the closest pairs, i.e., $D^{\min}_{\rm{ED}}=\min_{l\neq \hat{l}} D_{\rm{ED}}(l,\hat{l})$. Substituting (\ref{eq14}) into (\ref{eq13}), we have  
\begin{equation} \label{eq15}
\begin{split}
\gamma_G=&\frac{{r2^{r-1}}}{N_r({2^r-1})}+\\&\frac{10\log_{10}\frac{\lambda_{\textrm{B}}}{\lambda_{\textrm{D}}}\exp{\left\{D_{\rm{ED}}^{\min}(l,\hat{l})^2_{\textrm{D}}-D_{\rm{ED}}^{\min}(l,\hat{l})^2_{\textrm{B}}\right\}}}{N_r}~(\textrm{dB}),
\end{split}
\end{equation}
where the first term is the mapping gain  \cite{Guo2016};  the second term stands for the gain brought by the proposed signal shaping and reflecting methods; $\lambda_{\textrm{B}}$, $\lambda_{\textrm{D}}$ represent the numbers of closest pairs in the system with DJMSR and the benchmark system; $D_{\rm{ED}}^{\min}(l,\hat{l})^2_{\textrm{D}}$, $D_{\rm{ED}}^{\min}(l,\hat{l})^2_{\textrm{B}}$ refer to the Euclidean distance of the closest pairs in the system with DJMSR and the benchmark system, respectively.

\section{Continuous Optimization-Based Joint Signal Mapping, Shaping and Reflecting Design (CJMSR)}
It is worth noting that the aforementioned phase-shift reflecting design in DJMSR are based on a given  $\Psi$, which corresponds to the RIS equipped with discrete phase-shift reflecting units or ON/OFF reflecting units. With a large number of continuous phase shift units equipped on the RIS, the set size of the feasible reflecting patterns $\Psi$ goes to infinity. The discrete optimization-based reflecting design in DJMSR will fail. In addition, the sets $\mathcal{X}_1,\mathcal{X}_2,\cdots,\mathcal{X}_{K}$ in DJMSR are optimized with a given transmit signal set $\mathcal{X}$. Consequently, the performance is highly depending on $\mathcal{X}$. How to optimize $\mathcal{X}_1,\mathcal{X}_2,\cdots,\mathcal{X}_{K}$ in the generalized complex field instead of $\mathcal{X}$ remains unanswered. To address these issues, we will investigate the joint signal  shaping and reflecting design in the continuous fields. In our design, the optimization is firstly initialized with the output of DJMSR for JRM\&SRM. Then, we alternatively optimize the reflecting patterns and transmit signal sets to minimize the system BER.
\begin{figure*}
  \setcounter{equation}{29}
\begin{equation}\label{gra}
\begin{split}
{\nabla _{\textbf{q}} }g(\textbf{q}) =& - \frac{1}{Lr}\sum_{l=1}^{L}\sum_{\hat{l}=1,\hat{l}\neq l}^{L} D_{\rm{HD}}(\textbf{b}_l,\textbf{b}_{\hat{l}})\sqrt{\frac{1}{\pi\sigma^2 \left[\textbf{q}^H \textbf{C}_{l,\hat{l}}\textbf{q} + 2\Re\left(\textbf{q}^T{\textbf{a}_{l,\hat{l}}}\right) +\left\| {{\hat{\textbf{H}}_d}{{\hat{\textbf{x}}_{l,\hat{l}}}}} \right\|_2^2\right]}}\\&\times\exp \left( { - \frac{ \left[\textbf{q}^H \textbf{C}_{l,\hat{l}}\textbf{q} + 2\Re\left(\textbf{q}^T{\textbf{a}_{l,\hat{l}}}\right) +\left\| {{\hat{\textbf{H}}_d}{{\hat{\textbf{x}}_{l,\hat{l}}}}} \right\|_2^2\right]}{4\sigma^2}} \right)\left(\textbf{C}_{l,\hat{l}}\textbf{q}+\textbf{a}_{l,\hat{l}}^C\right)+\frac{{\left\| \textbf{q} \right\|_p^{1 - p}{\textbf{p}_{\textbf{q}} }}}{2t\left( { 1-{{\left\| \textbf{q}  \right\|}_p}} \right)}.
\end{split}
\end{equation}
\hrule
\end{figure*}

\subsection{Alternative Optimization}
\subsubsection{Continuous Optimization-Based Reflecting (COR) Design}
For either JRM or SRM,   given the non-empty transmit signal vector sets $\mathcal{X}_1,\mathcal{X}_2,\cdots, \mathcal{X}_{\tilde{K}}$ for $\tilde{K}$ reflecting patterns having been selected from a candidate set satisfying $\sum_{k=1}^{\tilde{K}}|\mathcal{X}_k|=L$ by DJMSR, optimizing multiple reflecting patterns is our new target. That is, we next aim to jointly optimize $\mathbf{\Phi}_1,\cdots, \mathbf{\Phi}_{\tilde{K}}$, to reflect the signals and simultaneously carry information.  The corresponding optimization problem can be formulated to be  
\setcounter{equation}{15}
\begin{equation}
\begin{split}
(\textbf{OP3}):~\mathrm{Given}: &~\textbf{H}_d,\textbf{H}_1,\textbf{H}_2,\sigma,\mathcal{X}_1,\mathcal{X}_2,\cdots, \mathcal{X}_{\tilde{K}}\\
\mathrm{Find}:&~\mathbf{\Phi}_1,\mathbf{\Phi}_2,\cdots, \mathbf{\Phi}_{\tilde{K}}\\
\mathrm{Minimize}:&~\overline{P_e}\\
\mathrm{Subject~to}:&~|(\mathbf{\Phi}_k)_{nn}|=1, 
(\mathbf{\Phi}_k)_{mn}=0, \forall m\neq n,\\
&~k=1,2,\cdots,{\tilde{K}}.
\end{split}
\end{equation}
To solve the problem, we introduce four new matrices 
\begin{equation}
\hat{\textbf{H}}_d=[\overbrace{\textbf{H}_d,\textbf{H}_d,\cdots,\textbf{H}_d}^{\tilde{K}}]\in\mathbb{C}^{N_r\times {\tilde{K}}N_t},
\end{equation}
\begin{equation}
\hat{\textbf{H}}_2=[\overbrace{\textbf{H}_2,\textbf{H}_2,\cdots,\textbf{H}_2}^{\tilde{K}}]\in\mathbb{C}^{N_r\times {\tilde{K}}N},
\end{equation}
\begin{equation}
\hat{\textbf{H}}_1=
\begin{bmatrix}
    ~\textbf{H}_1 & \textbf{0} &\cdots&\textbf{0}\\
    ~\textbf{0} & ~\textbf{H}_1& \cdots& \textbf{0}~ \\
~\vdots& \vdots &  \ddots& \vdots~\\
    ~\textbf{0} &\textbf{0} &\cdots &\textbf{H}_1\\
\end{bmatrix}\in \mathbb{C}^{{\tilde{K}}N\times {\tilde{K}}N_t},
\end{equation} 
\begin{equation}
\textbf{D}_{\textbf{q}}=
\begin{bmatrix}
    ~\mathbf{\Phi}_1 & \textbf{0} &\cdots&\textbf{0}\\
    ~\textbf{0} & ~\mathbf{\Phi}_2& \cdots& \textbf{0}~ \\
~\vdots& \vdots &  \ddots& \vdots~\\
    ~\textbf{0} &\textbf{0} &\cdots &\mathbf{\Phi}_{\tilde{K}}\\
\end{bmatrix}\in \mathbb{C}^{{\tilde{K}}N\times {\tilde{K}}N},
\end{equation} 
and $L$ new vectors
\begin{equation}
\hat{\textbf{x}}_l=\textbf{g}_k\otimes\textbf{x}_i^k\in\mathbb{C}^{{\tilde{K}} N_t\times 1}, l=1,2,\cdots,L,
\end{equation}
where $\textbf{g}_k\in\mathbb{C}^{{\tilde{K}}\times 1}$ is the $k$th ${\tilde{K}}$-dimensional vector basis with all zeros except the $k$th entry being one. Moreover, we define $\textbf{q}=\diag\{\textbf{D}_q\}$. With these new matrices and vectors, the square of the Euclidean distances can be transformed to be
\begin{equation}\label{ED}
\begin{split}
D_{\rm{ED}}(l,\hat{l})^2=&||(\textbf{H}_d+\textbf{H}_2\mathbf{\Phi}_k\textbf{H}_1)\textbf{x}_i^k-(\textbf{H}_d+\textbf{H}_2\mathbf{\Phi}_{\hat{k}}\textbf{H}_1)\textbf{x}_{\hat{i}}^{\hat{k}}||_2^2\\
=&||(\hat{{\textbf{H}}}_d+\hat{\textbf{\textbf{H}}}_2\textbf{D}_{\textbf{q}}\hat{\textbf{H}}_1)(\hat{\textbf{x}}_l-\hat{\textbf{x}}_{\hat{l}})|| _2^2\\
=& \left\| {\hat{\textbf{\textbf{H}}}_2\textbf{D}_{\textbf{q}}\hat{\textbf{H}}_1{{\hat{\textbf{x}}_{l,\hat{l}}}}} \right\|_2^2+ {\hat{\textbf{x}}_{l,\hat{l}}}^H{\hat{{\textbf{H}}}_d^H}{\hat{\textbf{\textbf{H}}}_2\textbf{D}_{\textbf{q}}\hat{\textbf{H}}_1{{\hat{\textbf{x}}_{l,\hat{l}}}}}
\\&+\hat{\textbf{x}}_{l,\hat{l}}^H{\hat{\textbf{H}}_1^H}\textbf{D}_{\textbf{q}}^H{{\hat{\textbf{H}}_2^H} {\hat{\textbf{H}}_d}{{\hat{\textbf{x}}_{l,\hat{l}}}}}+\left\| {{\hat{\textbf{H}}_d}{{\hat{\textbf{x}}_{l,\hat{l}}}}} \right\|_2^2\\
=& \left\| {{\hat{\textbf{H}}_2}\textbf{D}_{\textbf{q}} {\hat{\textbf{H}}_1}{{\hat{\textbf{x}}_{l,\hat{l}}}}} \right\|_2^2+ 2\Re\left(\textbf{q}^T{\textbf{a}_{l,\hat{l}}}\right)+\left\| {{\hat{\textbf{H}}_d}{{\hat{\textbf{x}}_{l,\hat{l}}}}} \right\|_2^2,
\end{split}
\end{equation}
where ${\hat{\textbf{x}}_{l,\hat{l}}} = \hat{\textbf{x}}_l-\hat{\textbf{x}}_{\hat{l}}\in\mathbb{C}^{{\tilde{K}} N_t\times 1}$ and ${\textbf{a}_{l,\hat{l}}}\in\mathbb{C}^{{\tilde{K}} N\times 1}$ is a vector with $n$-th element ${(\textbf{a}_{{l,\hat{l}}})_n} = {{\textbf{x}_{l,\hat{l}}^H}{\hat{\textbf{h}}_{d,2,n}}\hat{\textbf{h}}_{1,n}^T{\textbf{x}_{l,\hat{l}}}}$, where ${\hat{\textbf{h}}_{d,2,n}}$ is the $n$-th column of ${\hat{\textbf{H}}_d^H}{{\hat{\textbf{H}}_2}}$ and $\hat{\textbf{h}}_{1,n}$ denotes the $n$-th column of $\hat{\textbf{H}}_{1}^T$. 
The first term of (\ref{ED}) can be written as
\begin{align}\label{sEd}
{\left\| {{\hat{\textbf{H}}_2}\textbf{D}_{\textbf{q}}\hat{\textbf{H}}_1{{\hat{\textbf{x}}_{l,\hat{l}}}}} \right\|_2^2}&={\textbf{m}_{l,\hat{l}}^H}\textbf{D}_{\textbf{q}}^H{\hat{\textbf{H}}_2}^H{\hat{\textbf{H}}_2}\textbf{D}_{\textbf{q}}{\textbf{m}_{l,\hat{l}}}\notag\\
&= \textrm{tr}\left(\textbf{D}_{\textbf{q}}^H{\textbf{R}_{\hat{\textbf{H}}_2}}\textbf{D}_{\textbf{q}}\Delta {\textbf{M}_{l,\hat{l}}}\right),
\end{align}
where ${\textbf{m}_{l,\hat{l}}} = {\hat{\textbf{H}}_1}{{\hat{\textbf{x}}_{l,\hat{l}}}}\in\mathbb{C}^{{\tilde{K}} N\times 1}$, $\textbf{R}_{\hat{\textbf{H}}2} = {\hat{\textbf{H}}_2}^H{\hat{\textbf{H}}_2}\in\mathbb{C}^{{\tilde{K}} N\times {\tilde{K}} N}$ and $\Delta {\textbf{M}_{l,\hat{l}}} = {{\textbf{m}_{l,\hat{l}}}}{{\textbf{m}_{l,\hat{l}}^H}}\in\mathbb{C}^{{\tilde{K}} N\times {\tilde{K}} N}$. Based on the equation $\textrm{tr}\left( {\textbf{D}_\textbf{u}^H\textbf{A}{\textbf{D}_\textbf{v}}{\textbf{B}^T}} \right) = {\textbf{u}^H}\left( {\textbf{A} \odot \textbf{B}} \right)\textbf{v}$ with $\textbf{D}_\textbf{u} = \rm{diag}\left\{\textbf{u}\right\}$ and $\textbf{D}_\textbf{v} = \rm{diag}\left\{\textbf{v}\right\}$ \cite{Guo2019b}, the term ${\left\| {{\hat{\textbf{H}}_2}\textbf{D}_{\textbf{q}}\hat{\textbf{H}}_1{{\hat{\textbf{x}}_{l,\hat{l}}}}}\right\|_2^2}$ can be transformed to be
\begin{align}\label{pphi}
{\left\| {{\hat{\textbf{H}}_2}\textbf{D}_{\textbf{q}}\hat{\textbf{H}}_1{{\hat{\textbf{x}}_{l,\hat{l}}}}}\right\|_2^2} &=\textbf{q}^H\left(\textbf{R}_{\textbf{H}2}\odot\Delta {\textbf{M}_{l,\hat{l}}^T}\right)\textbf{q}\notag\\
& =\textbf{q}^H \textbf{C}_{l,\hat{l}}\textbf{q},
\end{align}
where $\textbf{C}_{l,\hat{l}} = \textbf{R}_{\hat{\textbf{H}}2}\odot\Delta {\textbf{M}_{l,\hat{l}}^T}\in\mathbb{C}^{{\tilde{K}} N\times {\tilde{K}} N}$. Submitting the term into (\ref{ED}) yields
\begin{align}\label{dis}
D_{\rm{ED}}(l,\hat{l})^2 &=\textbf{q}^H \textbf{C}_{l,\hat{l}}\textbf{q} + 2\Re\left(\textbf{q}^T{\textbf{a}_{l,\hat{l}}}\right) +\left\| {{\hat{\textbf{H}}_d}{{\hat{\textbf{x}}_{l,\hat{l}}}}} \right\|_2^2.
\end{align}
The objective function of (\textbf{OP3}) becomes
\begin{equation}
\begin{split}
\overline{P_e}(\textbf{q})=&\frac{1}{Lr}\sum_{l=1}^{L}\sum_{\hat{l}=1,\hat{l}\neq l}^{L}D_{\rm{HD}}(\textbf{b}_l,\textbf{b}_{\hat{l}})\times \\&Q\left(\sqrt{\frac{\textbf{q}^H \textbf{C}_{l,\hat{l}}\textbf{q} + 2\Re\left(\textbf{q}^T{\textbf{a}_{l,\hat{l}}}\right) +\left\| {{\hat{\textbf{H}}_d}{{\hat{\textbf{x}}_{l,\hat{l}}}}} \right\|_2^2}{2\sigma^2}}\right).
\end{split}
\end{equation}
Then, the problem (\textbf{OP3}) is transformed to be
\begin{equation}
\begin{split}
(\textbf{OP4}):~\mathrm{Given}: &~\textbf{C}_{l,\hat{l}}, l\neq \hat{l}=1,2,\cdots,L\\
\mathrm{Find}:&~\textbf{q}\\
\mathrm{Minimize}:&~\overline{P_e}\\
\mathrm{Subject~to}:&~|q_n|=1,n=1,2,\cdots,\tilde{K}N.
\end{split}
\end{equation}
Similarly to \cite{Ye2019}, we relax the unit modulus constraint to be
$\text{tr}\left( \textbf{q}\textbf{q}^H \right) = \tilde{K}N$ and ${\left\| \textbf{q}  \right\|_\infty } \le 1$. To deal with constraint ${\left\| \textbf{q}  \right\|_\infty } \le 1$, we use a large $p$ norm to replace the infinity norm and rebuild a new objective function as
\begin{equation}\label{gq}
g(\textbf{q})=\overline{P_e}(\textbf{q})+ B(1-||\textbf{q}||_p),
\end{equation}
where $B(\cdot)$ is a penalty function defined by
\begin{align}
B\left( u \right) = \left\{ {\begin{array}{*{20}{c}}
{ - \frac{1}{t}\ln \left( u \right),}&{u > 0}\\
{\infty ,}&{u \le 0,}
\end{array}} \right.
\end{align}
and $t$ is the penalty parameter. Taking the first derivation of the objective function $g(\textbf{q})$ with respect to $\textbf{q}$ yields (\ref{gra}), where $\textbf{p}_{\textbf{q}}={\left[ {{q _1} \cdot {{\left| {{q _1}} \right|}^{p - 2}},{q _2} \cdot {{\left| {{q _2}} \right|}^{p - 2}},...,{q  _{\tilde{K}N}} \cdot {{\left| {{q _{\tilde{K}N}}} \right|}^{p - 2}}} \right]^T}$.

To satisfy the constraint $\text{tr}\left( \textbf{q}\textbf{q}^H \right) = \tilde{K}N$, we perform a projection as
 \setcounter{equation}{30}
\begin{equation}\label{PJ}
\Delta \textbf{q}=-\textbf{P}_{1}{\nabla _{\textbf{q}}}g(\textbf{q})
\end{equation}
to ensure $\textbf{q}^H\Delta \textbf{q}=0$ in the update process, where $\textbf{P}_1$ is a projection matrix  given by
$\textbf{P}_{1}=\textbf{I}_{\tilde{K}N}-\frac{\textbf{q}\textbf{q}^H}{\tilde{K}N}$ and $\textbf{P}_{1} = \textbf{P}_{1}^H= \textbf{P}_{1}^2=\textbf{P}_{1}^H\textbf{P}_{1}$. Based on the obtained $\Delta \textbf{q}$, we can search the solution iteratively as listed in Algorithm 3.
\begin{proposition}
Algorithm 3 converges.
\end{proposition}
\begin{IEEEproof}
During each iteration in Algorithm 3, the variance of the objective function in the update process from $\textbf{q}_{\textrm{iter}} $ to $\textbf{q}_{\textrm{iter}+1}$ can be expressed as
\begin{equation}
\begin{split}
g(\textbf{q}_{\textrm{iter}+1})-g(\textbf{q}_{\textrm{iter}})&=g(\textbf{q}_{\textrm{iter}} + \alpha_{\textrm{iter}} \Delta \textbf{q}_{\textrm{iter}})-g(\textbf{q}_{\textrm{iter}})\\
&\leq u\alpha_{\textrm{iter}}[{\nabla _{\textbf{q}} }g(\textbf{q})]^H \Delta\textbf{q}_{\textrm{iter}}\\
&=-u\alpha_{\textrm{iter}}[{\nabla _{\textbf{q}} }g(\textbf{q})]^H \textbf{P}_{1}{\nabla _{\textbf{q}}}g(\textbf{q})
\end{split}
\end{equation}
where $u$ is a small real positive number in the Wolfe conditions \cite{Boyd2004} and $\alpha_{\textrm{iter}}>0$.
Substituting $\textbf{P}_{1}=\textbf{P}_{1}^H\textbf{P}_{1}$, we have 
\begin{equation}
g(\textbf{q}_{\textrm{iter}+1})-g(\textbf{q}_{\textrm{iter}})\leq -c\alpha_{\textrm{iter}}|| \textbf{P}_{1}{\nabla _{\textbf{q}}}g(\textbf{q})||_2^2<0.
\end{equation}
Since $g(\textbf{q}_{\textrm{iter}+1})$ in (\ref{gq}) is lower bounded, Algorithm 1 will converge.
\end{IEEEproof}

\begin{algorithm}[t]
\caption{Continuous-Optimization Based Reflecting Design}
\label{10}
\begin{algorithmic}[1]
\STATE {Initialize a feasible solution ${\textbf{q}}_0$, a halting criterion ${\varepsilon_1}> 0$, a penalty parameter $t$, and the iteration number ${\textrm{iter}}=0$. }
\STATE {Compute the gradient according to (\ref{gra})}.
\STATE {Project the negative gradient to obtain $\Delta \textbf{q}$ according to (\ref{PJ})}.
\STATE Update
$\textbf{q}_{\textrm{iter}+1} = \textbf{q}_{\textrm{iter}} + \alpha_{\textrm{iter}} \Delta \textbf{q}_{\textrm{iter}}$,
where $ \alpha_{\textrm{iter}}$ is computed by a line search procedure to satisfy the Wolfe conditions \cite{Boyd2004} and $\textrm{iter}\leftarrow\textrm{iter}+1$.
\STATE Repeat steps 2-4 until the halting criterion $[g(\textbf{q}_{\textrm{iter}})-g(\textbf{q}_{\textrm{iter+1}})]/g(\textbf{q}_{\textrm{iter}})\le \varepsilon_1$ is met.
\STATE Output $q_n^*=\frac{q_n}{|q_n|}$ for all $n=1,2,\cdots,\tilde{K}N$ and $\mathbf{\Phi}_k=\diag([q_{(k-1)N+1}^*,\cdots, q_{kN+1}^*]^T), k=1,2,\cdots,\tilde{K}$.
\end{algorithmic}
\end{algorithm}

\subsubsection{Continuous Optimization-Based Signal Shaping (COS) Design}
For JRM, $\mathcal{X}_1,,\cdots, \mathcal{X}_{\tilde{K}}$ may be different. Based on  $\mathbf{\Phi}_1,\cdots, \mathbf{\Phi}_{\tilde{K}}$  and the set sizes $|\mathcal{X}_1|,,\cdots, |\mathcal{X}_{\tilde{K}}|$, set entries of $\mathcal{X}_1,\cdots, \mathcal{X}_{\tilde{K}}$ can be further optimized in the complex field by solving the following problem:
\begin{equation}
\begin{split}
(\textbf{OP5}):~\mathrm{Given}: &~\textbf{H}_d,\textbf{H}_1,\textbf{H}_2,\sigma,\mathbf{\Phi}_1,\mathbf{\Phi}_2,\cdots, \mathbf{\Phi}_{\tilde{K}},\\&~|\mathcal{X}_1|,|\mathcal{X}_2|,\cdots, |\mathcal{X}_{\tilde{K}}|, L\\
\mathrm{Find}:&~\mathcal{X}_1,\mathcal{X}_2,\cdots, \mathcal{X}_{\tilde{K}}\\
\mathrm{Minimize}:&~\overline{P_e}\\
\mathrm{Subject~to}:&~\frac{1}{L}\sum_{k=1}^{K}\sum_{\textbf{x}_i^k\in\mathcal{X}_k}||\textbf{x}_i^k||^2_2=1.
\end{split}
\end{equation}
Writing $\textbf{G}_k=\textbf{H}+\textbf{H}_2\mathbf{\Phi}_k\textbf{H}_1$, we re-express the square of the Euclidean distances as
\begin{equation}\label{ED1}
\begin{split}
D_{\rm{ED}}(l,\hat{l})^2=&||\textbf{G}_k\textbf{x}_i^k-\textbf{G}_{\hat{k}}\textbf{x}_{\hat{i}}^{\hat{k}}||_2^2.
\end{split}
\end{equation}
For neatness, we introduce several new matrices
\begin{equation}
\begin{split}
\textbf{W}&=\left[\overbrace{\textbf{G}_1,\cdots,\textbf{G}_1}^{|\mathcal{X}_1|},\overbrace{\textbf{G}_2,\cdots,\textbf{G}_2,}^{|\mathcal{X}_2|},\cdots,\overbrace{\textbf{G}_{\tilde{K}},\cdots,\textbf{G}_{\tilde{K}}}^{|\mathcal{X}_{\tilde{K}}|}\right]\\&\in\mathbb{C}^{N_r\times LN_t},
\end{split}
\end{equation}
\begin{equation}
\begin{split}
\textbf{D}_{\textbf{z}}&\triangleq 
\begin{bmatrix}
    ~\textbf{X}_1^1 & \textbf{0} & \textbf{0}&\cdots&\textbf{0}&\textbf{0}&  \textbf{0}~ \\
    ~\textbf{0} & \ddots& \textbf{0}& \cdots&\cdot&\textbf{0}& \textbf{0}~ \\
     ~\textbf{0} & \textbf{0} & \textbf{X}_{|\mathcal{X}_1|}^1&  \cdots&\textbf{0}&\cdot&\textbf{0}~\\
~\vdots& \vdots & \vdots& \ddots& \vdots&\vdots&\vdots~\\
    ~\textbf{0} & \cdot & \textbf{0} &\cdots &\textbf{X}_1^K& \textbf{0}&  \textbf{0}~\\
 ~\textbf{0} & \textbf{0} & \cdot &\cdots &\textbf{0}& \ddots&  \textbf{0}~\\
 ~\textbf{0} & \textbf{0} & \textbf{0} &\cdots &\textbf{0}& \textbf{0}&  \textbf{X}_{|\mathcal{X}_{\tilde{K}}|}^K~
\end{bmatrix}\\&\in\mathbb{C}^{LN_t\times LN_t},
\end{split}
\end{equation}
where
\begin{equation}
\textbf{X}_i^k=\diag{(\textbf{x}_i^k)}\in\mathbb{C}^{N_t\times N_t}.
\end{equation}
We define $\textbf{z}=\diag\{\textbf{D}_{\textbf{z}}\}\in\mathbb{C}^{LN_t\times 1}$ and  $\textbf{o}_l=\textbf{e}_l\otimes \textbf{1}_{N_{t}\times 1}\in\mathbb{C}^{LN_t\times 1}$, $l=1,2,\cdots, L$,  where $\textbf{e}_l$ is the $l$th $L$-dimensional vector basis with all zeros except the $l$th entry being one. 

Based on these new matrices and vectors, the square of the pairwise Euclidean distances can be expressed as
\begin{equation}\label{eqD}
\begin{split}
D_{\rm{ED}}(l,\hat{l})^2&=||\textbf{G}_k\textbf{x}_i^k-\textbf{G}_{\hat{k}}\textbf{x}_{\hat{i}}^{\hat{k}}||_2^2\\&=||\textbf{W}\textbf{D}_{\textbf{z}}\textbf{o}_l-\textbf{W}\textbf{D}_{\textbf{z}}\textbf{o}_{\hat{l}}||_2^2\\
&=(\textbf{o}_l-\textbf{o}_{\hat{l}})^H\textbf{D}_\textbf{z}^H\textbf{W}^H\textbf{W}\textbf{D}_\textbf{z}(\textbf{o}_l-\textbf{o}_{\hat{l}})\\
&=\tr\left(\textbf{D}_\textbf{z}^H\textbf{R}_{\textbf{W}}\textbf{D}_\textbf{z}\Delta\textbf{O}_{l{\hat{l}}}\right)\\
&\overset{(a)}=\textbf{z}^H\textbf{Z}_{l,\hat{l}}\textbf{z},
\end{split}
\end{equation}
where $\textbf{R}_{\textbf{W}}=\textbf{W}^H\textbf{W}\in\mathbb{C}^{LN_t\times LN_t}$, $\Delta\textbf{O}_{l{\hat{l}}}=(\textbf{o}_l-\textbf{o}_{\hat{l}})(\textbf{o}_l-\textbf{o}_{\hat{l}})^H\in\mathbb{C}^{LN_t\times LN_t}$ and $\textbf{Z}_{l,\hat{l}}=\textbf{R}_{\textbf{H}\textbf{G}}\odot\Delta\textbf{O}_{l{\hat{l}}}^H\in \mathbb{C}^{LN_t\times LN_t}$.
The equality $(a)$ holds due to 
 $\tr(\textbf{D}_{\textbf{u}}\textbf{U}\textbf{D}_{\textbf{v}}\textbf{V}^H)=\textbf{u}^H(\textbf{U}\odot\textbf{V})\textbf{v}$ for any two diagonal matrices $\textbf{D}_{\textbf{v}}=\diag(\textbf{v})$ and $\textbf{D}_{\textbf{u}}=\diag(\textbf{u})$. 
Based on the reformulation, (\textbf{OP5}) becomes
\begin{equation}
\begin{split}
(\textbf{OP6}):~\mathrm{Given}: &~\textbf{Z}_{l,\hat{l}}, l\neq \hat{l}=1,2,\cdots,L\\
\mathrm{Find}:&~\textbf{z}\\
\mathrm{Minimize}:&~\overline{P_e}(\textbf{z})\\
\mathrm{Subject~to}:&~\text{tr}\left( \textbf{z}\textbf{z}^H \right) = L.
\end{split}
\end{equation}
where 
\begin{equation}
\begin{split}
\overline{P_e}(\textbf{z})=&\frac{1}{Lr}\sum_{l=1}^{L}\sum_{\hat{l}=1,\hat{l}\neq l}^{L}D_{\rm{HD}}(\textbf{b}_l,\textbf{b}_{\hat{l}}) Q\left(\sqrt{\frac{\textbf{z}^H\textbf{Z}_{l,\hat{l}}\textbf{z}}{2\sigma^2}}\right).
\end{split}
\end{equation}
Taking the first derivation of the objective function $\overline{P_e}(\textbf{z})$ with respect to $\textbf{z}$ yields
\begin{equation}\label{graz}
\begin{split}
{\nabla _{\textbf{z}} }\overline{P_e}(\textbf{z}) =& - \frac{1}{Lr}\sum_{l=1}^{L}\sum_{\hat{l}=1,\hat{l}\neq l}^{L} D_{\rm{HD}}(\textbf{b}_l,\textbf{b}_{\hat{l}})\sqrt{\frac{1}{\pi\sigma^2 \textbf{z}^H\textbf{Z}_{l,\hat{l}}\textbf{z}}}\times\\&\exp \left( { - \frac{ \textbf{z}^H\textbf{Z}_{l,\hat{l}}\textbf{z}}{4\sigma^2}} \right)\textbf{Z}_{l,\hat{l}}\textbf{z},
\end{split}
\end{equation}
To satisfy the constraint $\text{tr}\left( \textbf{z}\textbf{z}^H \right) = L$, we perform a projection
\begin{equation}\label{PJz}
\Delta \textbf{z}=-\textbf{P}_{2}{\nabla _{\textbf{z}} }\overline{P_e}(\textbf{z})
\end{equation}
to ensure $\textbf{z}^H\Delta \textbf{z}=0$,
where $\textbf{P}_{2}$ is a projection matrix given by
$\textbf{P}_{2}=\textbf{I}_{L}-\frac{\textbf{z}\textbf{z}^H}{L}$ satisfying $\textbf{P}_{2} = \textbf{P}_{2}^H= \textbf{P}_{2}^2=\textbf{P}_{2}^H\textbf{P}_{2}$. Based on the derived search direction, the continuous optimization algorithm can be conducted as listed in Algorithm 4.

\begin{algorithm}[t]
\caption{Continuous Optimization-Based Signal Shaping for JRM}
\label{10}
\begin{algorithmic}[1]
\STATE {Initialize a feasible solution ${\textbf{z}}_0$, a halting criterion ${\varepsilon_2}> 0$, and the iteration number ${\textrm{iter}}=0$. }
\STATE {Compute the gradient according to (\ref{graz})}.
\STATE {Project the negative gradient to obtain $\Delta \textbf{z}$ according to (\ref{PJz})}.
\STATE Update
$\textbf{z}_{\textrm{iter}+1} = \textbf{z}_{\textrm{iter}} + \beta_{\textrm{iter}} \Delta \textbf{z}_{\textrm{iter}}$,
where $ \beta_{\textrm{iter}}$ is computed by a line search procedure to satisfy the Wolfe conditions \cite{Boyd2004} and $\textrm{iter}\leftarrow\textrm{iter}+1$.
\STATE Repeat steps 2-4 until the halting criterion $[\overline{P_e}(\textbf{z}_{\textrm{iter}})-\overline{P_e}(\textbf{z}_{\textrm{iter+1}})]/\overline{P_e}(\textbf{z}_{\textrm{iter}})\le \varepsilon_2$ is met.
\STATE Output $\textbf{x}_i^k=\hat{\textbf{W}}\textbf{D}_{\textbf{z}}\textbf{o}_l$, where is defined by $\hat{\textbf{W}}=[\overbrace{\textbf{I}_{N_t},\cdots,\textbf{I}_{N_t}}^L]\in\mathbb{C}^{N_t\times LN_t}$.
\end{algorithmic}
\end{algorithm}

For SRM, $\tilde{K}=K_c$. Owing to $\mathcal{X}_1=\mathcal{X}_2=\cdots=\mathcal{X}_{\tilde{K}}$, the optimization problem becomes 
\begin{equation}
\begin{split}
(\textbf{OP7}):~\mathrm{Given}: &~\textbf{G}_1,\textbf{G}_2,\cdots, \textbf{G}_{\tilde{K}},\sigma,M_c\\
\mathrm{Find}:&~\mathcal{X}_1,\mathcal{X}_2,\cdots, \mathcal{X}_{\tilde{K}}\\
\mathrm{Minimize}:&~\overline{P_e}\\
\mathrm{Subject~to}:&~\frac{1}{M_c}\sum_{\textbf{x}_i\in\mathcal{X}_1}||\textbf{x}_i||^2_2=1,\\
&~\mathcal{X}_1=\mathcal{X}_2=\cdots=\mathcal{X}_{\tilde{K}},\\
&~|\mathcal{X}_1|=|\mathcal{X}_2|=\cdots=|\mathcal{X}_{\tilde{K}}|=M_c.
\end{split}
\end{equation}
By introducing $\textbf{c}=[\textbf{x}_1^T,\textbf{x}_2^T,\cdots,\textbf{x}_{M_c}^T]^T\in\mathbb{C}^{M_cN_t\times 1}$, $\textbf{f}=[\Re{(\textbf{c}})^T,\Im{(\textbf{c}})^T]^T\in\mathbb{R}^{2M_cN_t\times 1}$, we transform (\textbf{OP7}) to be
\begin{equation}
\begin{split}
(\textbf{OP8}):~\mathrm{Given}: &~\textbf{G}_1,\textbf{G}_2,\cdots, \textbf{G}_{\tilde{K}},\sigma,M_c\\
\mathrm{Find}:&~\textbf{f}\\
\mathrm{Minimize}:&~\overline{P_e}(\textbf{f})\\
\mathrm{Subject~to}:&~\text{tr}\left( \textbf{f}\textbf{f}^H \right) = M_c.
\end{split}
\end{equation}
To solve it, we compute the first derivative of $\overline{P_e}(\textbf{f})$ with respect to $\textbf{f}$ to be
\begin{equation}\label{gradf}
\nabla_{\textbf{f}} \overline{P_e}(\textbf{f})=[\nabla_{f_1} \overline{P_e}(\textbf{f}),\nabla_{f_2} \overline{P_e}(\textbf{f}),\cdots,\nabla_{f_{2M_cN_t}} \overline{P_e}(\textbf{f}_{2M_cN_t})]^T,
\end{equation}
where $\nabla_{f_i} \overline{P_e}(\textbf{f})$ is approximately computed by
\begin{equation}
\nabla_{f_i} \overline{P_e}(\textbf{f})\approx \frac{\overline{P_e}(\textbf{f}+\delta \textbf{f}_i)-\overline{P_e}(\textbf{f})}{\delta},
\end{equation}
with $\delta$ being a small number and $\textbf{f}_i$ denoting the $i$-th $2M_cN_t\times 1$ base vector.  To satisfy the constraint $\text{tr}\left( \textbf{f}\textbf{f}^H \right) = M_c$, we perform a projection as
\begin{equation}\label{PJf}
\Delta \textbf{z}=-\textbf{P}_{3}\nabla_{\textbf{f}} \overline{P_e}(\textbf{f}),
\end{equation}
to ensure $\textbf{f}^H\Delta \textbf{f}=0$, where $\textbf{P}_3$ is the projection matrix given by
$\textbf{P}_{3}=\textbf{I}_{M_c}-\frac{\textbf{q}\textbf{q}^H}{M_c}$ satisfying $\textbf{P}_{3} = \textbf{P}_{3}^H= \textbf{P}_{3}^2=\textbf{P}_{3}^H\textbf{P}_{3}$. Based on the derived search direction, the continuous optimization algorithm can be conducted as listed in Algorithm 5.
\begin{proposition}
Both Algorithms 4 and 5 converge.
\end{proposition}
\begin{IEEEproof}
This proposition can be proved similarly to Proposition 1.
\end{IEEEproof}

\begin{algorithm}[t]
\caption{Continuous Optimization-Based Signal Shaping for SRM}
\label{10}
\begin{algorithmic}[1]
\STATE {Initialize a feasible solution ${\textbf{f}}_0$, a halting criterion ${\varepsilon_3}> 0$, and the iteration number ${\textrm{iter}}=0$. }
\STATE {Compute the gradient according to (\ref{gradf})}.
\STATE {Project the negative gradient to obtain $\Delta \textbf{z}$ according to (\ref{PJf})}.
\STATE Update
$\textbf{f}_{\textrm{iter}+1} = \textbf{f}_{\textrm{iter}} + \eta_{\textrm{iter}} \Delta \textbf{f}_{\textrm{iter}}$,
where $ \eta_{\textrm{iter}}$ is computed by a line search procedure to satisfy the Wolfe conditions \cite{Boyd2004} and $\textrm{iter}\leftarrow\textrm{iter}+1$.
\STATE Repeat steps 2-4 until the halting criterion $[\overline{P_e}(\textbf{f}_{\textrm{iter}})-\overline{P_e}(\textbf{f}_{\textrm{iter+1}})]/\overline{P_e}(\textbf{f}_{\textrm{iter}})\le \varepsilon_3$ is met.
\STATE Compute $\textbf{c}=[f_{1},\cdots f_{M_cN_t}]^T+j[f_{M_cN_t+1},\cdots f_{2M_cN_t}]^T$.
\STATE Output $\textbf{x}_i=[c_{(i-1)N_t+1},c_2,\cdots,c_{iN_t}]^T$.
\end{algorithmic}
\end{algorithm}

Based on the proposed COR and COS designs, the reflecting and signal shaping can be alternatively optimized. We summarize the CJMSR  in Algorithm 6. Propositions 1 and 2  can guarantee the convergence of Algorithm 6.

\begin{algorithm}[t]
\caption{CJMSR Design for JRM\&SRM}
\label{10}
\begin{algorithmic}[1]
\STATE {\textbf{Input:} $\textbf{H}_d,\textbf{H}_1,\textbf{H}_2,\sigma,\mathcal{X},\Psi, L, K_c, M_c$. }
\STATE{Perform DJMSR to obtain $\mathcal{X}_1,\mathcal{X}_2,\cdots,\mathcal{X}_{\tilde{K}}$.}
\STATE{Update $\mathbf{\Phi}_1,\mathbf{\Phi}_2,\cdots,\mathbf{\Phi}_{\tilde{K}}$ by performing COR design in use of Algorithm 3.}
\STATE{Update $\mathcal{X}_1,\mathcal{X}_2,\cdots,\mathcal{X}_{\tilde{K}}$ by performing the COS design in use of Algorithm 4 or 5.}
\STATE{Repeat step 2 and 3 until the halting criterion is met.}
\STATE{Update the mapping rules by performing the BSA \cite{Guo2016}.}
\end{algorithmic}
\end{algorithm}

\subsection{Computational Complexity and Performance Analysis}

In COR, the gradient calculation dominates the computational complexity, involving calculating $\textbf{C}_{l,\hat{l}} $ and $\textbf{q}^H \textbf{C}_{l,\hat{l}} \textbf{q}$ for all $l\neq l'$. Based on the fact, the complexity order of COR can be analyzed to be
\begin{align}
\mathcal{C}_{\textrm{COR}}=\mathcal{O}\left[ L^2\tilde{K}^2N^2(N_r+N_t)N_{\textrm{iter3}}\right],
\end{align}
where $N_{\textrm{iter3}}$ represents the number of iterations in Algorithm 3.

In COS for JRM,   the computation of $\textbf{Z}_{l,\hat{l}} $ and $\textbf{z}^H \textbf{Z}_{l,\hat{l}} \textbf{z}$ are needed, which introduces the complexity:
\begin{equation}
\mathcal{C}_{\textrm{COS}}^{\textrm{JRM}}=\mathcal{O}\left[ L^2(L^2N_t^2+\tilde{K}N)(N_r+N_t)N_{\textrm{iter4}}\right],
\end{equation}
where $N_{\textrm{iter4}}$ stands for the number of iterations in Algorithm 4.

In COS for SRM,   the computation of the $\overline{P_e}$ have to be computed $2M_cN_t+1$ times, which introduces the complexity: 
\begin{equation}
\mathcal{C}_{\textrm{COS}}^{\textrm{SRM}}=\mathcal{O}\left[ M_cN_tLN^2(N_r+N_t)N_{\textrm{iter5}}+L^2N_rN_{\textrm{iter5}}\right],
\end{equation}
where $N_{\textrm{iter5}}$ denotes for the number of iterations in Algorithm 5.

Thus, the aggregate computational complexity of CJMSR for JRM\&SRM can be computed by
\begin{equation}
\mathcal{C}_{\textrm{CJMSR}}^{\textrm{JRM}}=\mathcal{C}_{\textrm{DJMSR}}^{\textrm{JRM}}+N_{\textrm{iter6}}^{\textrm{J}}\left(\mathcal{C}_{\textrm{COR}}+\mathcal{C}_{\textrm{COS}}^{\textrm{JRM}}\right),
\end{equation} 
and
\begin{equation}
\mathcal{C}_{\textrm{CJMSR}}^{\textrm{SRM}}=\mathcal{C}_{\textrm{DJMSR}}^{\textrm{SRM}}+N_{\textrm{iter6}}^{\textrm{S}}\left(\mathcal{C}_{\textrm{COR}}+\mathcal{C}_{\textrm{COS}}^{\textrm{SRM}}\right),
\end{equation} 
where $N_{\textrm{iter6}}^{\textrm{J}}$ and $N_{\textrm{iter6}}^{\textrm{S}}$ refers to the numbers of iterations in use of Algorithm 6 for CJMSR for JRM\&SRM, respectively.
Similarly to Section III-B, the performance improvement brought by the proposed CJMSR method can be quantified by the SNR gain. Moreover, we remark that the proposed signal shaping can be regarded as a joint optimization of the transmit beamforming and transmit symbol vectors, which inherently outperforms the transmit beamforming optimization with the given transmit symbol vectors, e.g., the work in \cite{Ye2019}.

\section{Numerical Results}

In this section, numerical results are presented to investigate the performance of the proposed JRM\&SRM in various $(N_t, N_r, N, r)$ RIS-based MIMO communication systems. It is divided by three subsections. Subsection V-A investigates the performance of the proposed JRM\&SRM with DJMSR based on given $\mathcal{X}$ and $\Psi$. The comparison among JRM, SRM, RIS-C, RIS-BC (e.g, RIS-SSK), RIS-SM
and PBIT is also presented. 
Subsection V-B studies the performance of the proposed JRM\&SRM with CJMSR.
Subsection V-C shows the system performance in the presence of channel estimation errors. All simulation results are evaluated over $1000$ Rayleigh channel realizations.

\subsection{Performance of JRM\&SRM with DJMSR}

\begin{figure}[t]
  \centering
  \includegraphics[width=0.5\textwidth]{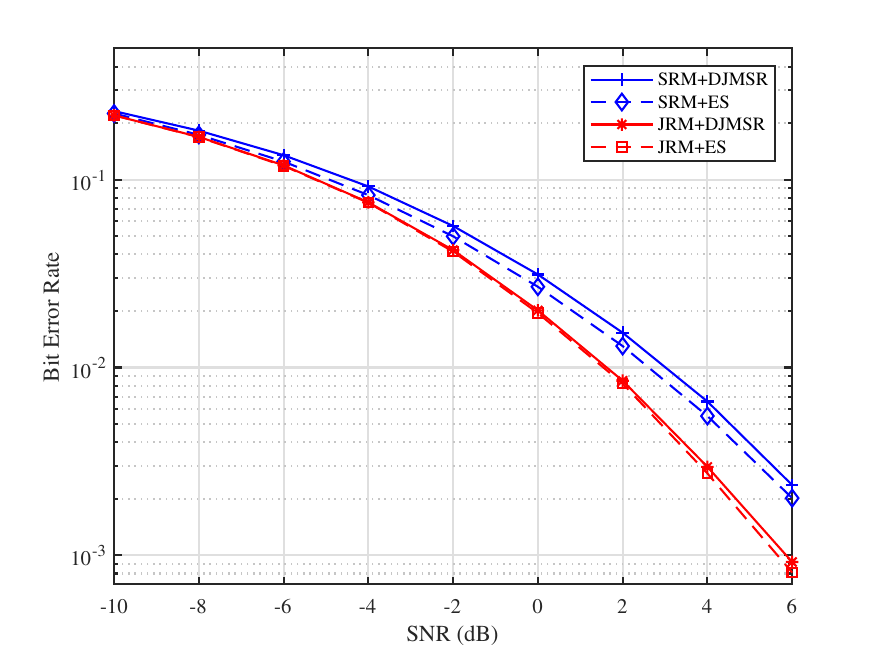}\\
 \caption{BER comparison between the proposed DJMSR and exhaustive search algorithms in $(1,3,4,3)$ RIS-based communication systems}
  \label{result1a}
\end{figure}
\begin{figure}[t]
  \centering
  \includegraphics[width=0.5\textwidth]{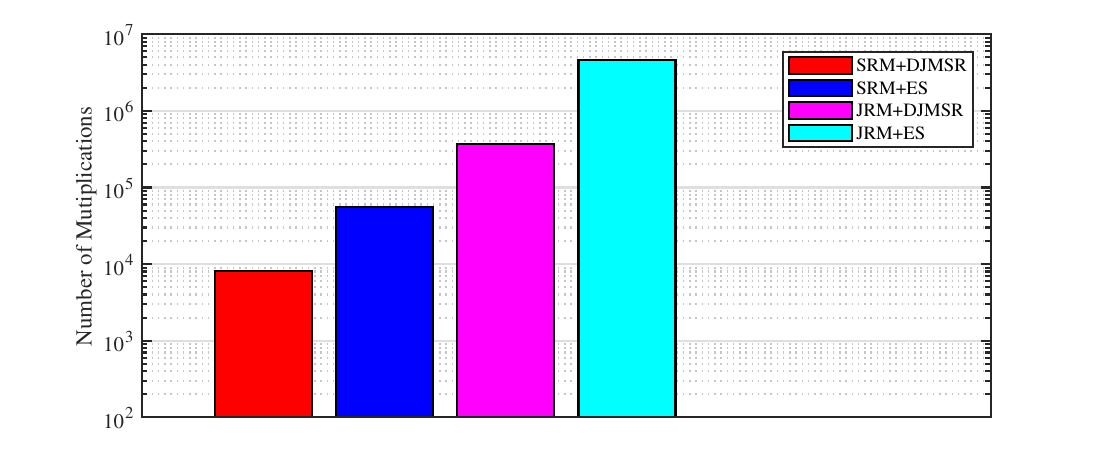}\\
 \caption{Computational complexity comparison of DJMSR with exhaustive search algorithm $(1,3,4,3)$ RIS-based communication systems in the number of multiplications.}
  \label{result1b}
\end{figure}

Firstly, we compare the proposed DJMSR for JRM\&SRM with  the exhaustive search (ES) algorithm in both performance and complexity in $(1,3,4,3)$ RIS-based communication systems. The numerical results regarding system BER and computational complexity are illustrated in Figs. \ref{result1a} and \ref{result1b},  where a randomly-generated  $\mathcal{X}$ with $|\mathcal{X}|=5$ and a randomly-generated $\Psi$ with $|\Psi|=3$ are adopted.  Results in Figs. \ref{result1a} and \ref{result1b} demonstrate that the proposed DJMSR for JRM can achieve almost the same performance as the ES solution, while the computational complexity is reduced by an order of magnitude in the number of multiplications.
It also demonstrated that the proposed DJMSR for SRM has a certain level of performance loss compared with the ES solution, but enjoys a considerably reduced complexity.

\begin{figure}[t]
  \centering
  \includegraphics[width=0.5\textwidth]{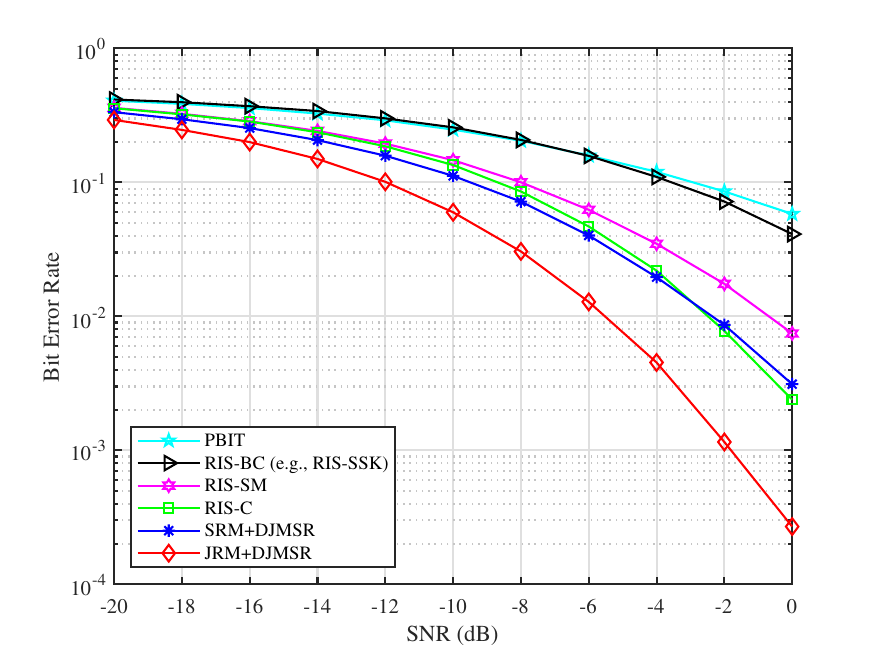}\\
 \caption{BER comparison among the proposed JRM\&SRM, existing RIS-C, RIS-BC (e.g., RIS-SSK), RIS-SM and PBIT in $(1,4,5,2)$ RIS-based communication systems.}
  \label{result2}
\end{figure}
As aforementioned in Section II, JRM\&SRM are more generalized and naturally superior to existing schemes. To verify this, we compare the proposed  JRM\&SRM with DJMSR with RIS-C, RIS-BC, RIS-SM and PBIT in a $(1,4,5,2)$ RIS-based communication system, as illustrated in Fig. \ref{result2}.
 To show the superiority of the proposed designs,
we use a union set $\mathcal{X}=\mathcal{X}_{\textrm{RIS-C}}\cup\mathcal{X}_{\textrm{RIS-BC}}\cup\mathcal{X}_{\textrm{RIS-SM}}\cup\mathcal{X}_{\textrm{PBIT}}$ as the signal candidate set and  a union set $\Psi=\Psi_{\textrm{RIS-C}}\cup\Psi_{\textrm{RIS-BC}}\cup\Psi_{\textrm{RIS-SM}}\cup\Psi_{\textrm{PBIT}}$ as the reflecting pattern candidate set.  Simulation results demonstrate that the optimized JRM\&SRM considerably outperform existing designs. In details, JRM with DJMSR outperforms RIS-C and RIS-SM by around 4-5 dB in the depicted high SNR regime. Compared to RIS-BC and PBIT, the performance gain becomes more obvious.  The performance differences mainly result from the bit mapping method, transmit signal shaping and reflecting patterns. Since RIS-C, RIS-BC, RIS-SM, and PBIT have additional constraints compared to RM, they are naturally less comparable.

\subsection{Performance of the Proposed JRM\&SRM with CJMSR}

\begin{figure}[t]
  \centering
  \includegraphics[width=0.5\textwidth]{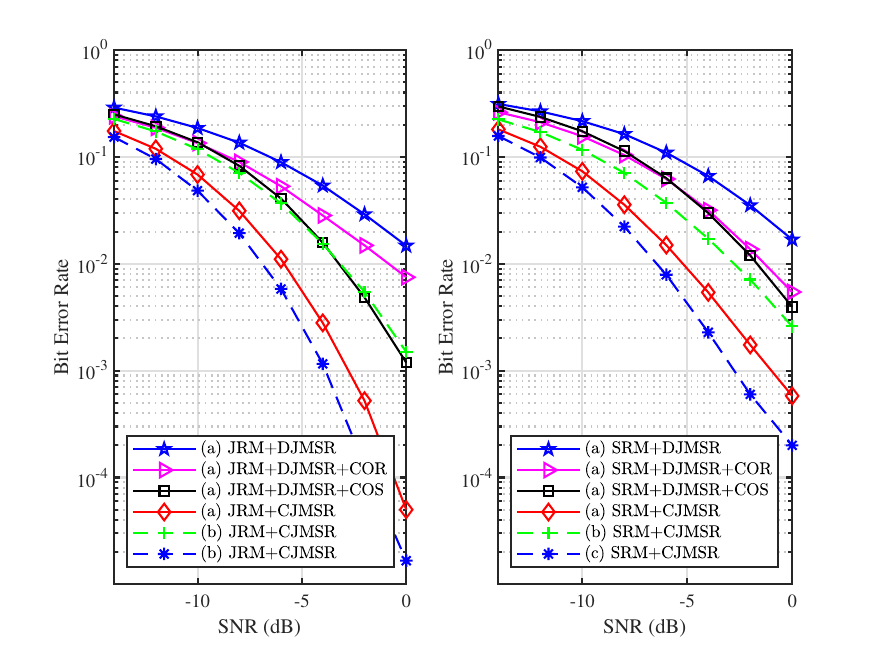}\\
 \caption{BER performance of the proposed JRM\&SRM with CJMSR in  (a):$(2, 3, 4, 3)$, (b): $(1, 3, 4, 3)$ and (c): $(2, 4, 4, 3)$ RIS-based MIMO systems.}
  \label{result3a}
\end{figure}

Secondly, we investigate the performance of CJMSR for JRM\&SRM in  (a):$(2, 3, 4, 3)$, (b): $(1, 3, 4, 3)$ and (c): $(2, 4, 4, 3)$ RIS-based MIMO systems, as illustrated in Fig. \ref{result3a}. In the simulations, we include the sole COR and COS for comparison. It is demonstrated that the sole COR brings around $1$ dB gain, the sole COS bring about $4$ dB gain, while CJMSR brings around $6$ dB in JRM systems. In SRM systems, the gains brought by COR, COS and CJMSR are $2$ dB, $2$ dB and $6$ dB, respectively. From these observations, we conclude that both COR and COS can considerably improve the performance and iteratively performing both in CJMSR enlarges the gain. Another significant insight gained from the results is that COS for JRM can provide more gain than COS for SRM. This is because that COS in JRM can optimize more optimization variables (i.e., $LN_t$ complex variables) than COS in SRM (i.e., $M_cN_t$ complex variables) and more optimization dimensions lead to a more considerable performance gain.  By comparing the performance of $(2, 3, 4, 3)$ systems with that of $(1, 3, 4, 3)$  and $(2, 4, 4, 3)$ systems, as expected, increasing the number of antennas either at the transmitter side  or at the receiver side brings better performance.

\subsection{Performance of the Proposed JRM With the Increase of the Number of RIS Units}
\begin{figure}[t]
  \centering
  \includegraphics[width=0.5\textwidth]{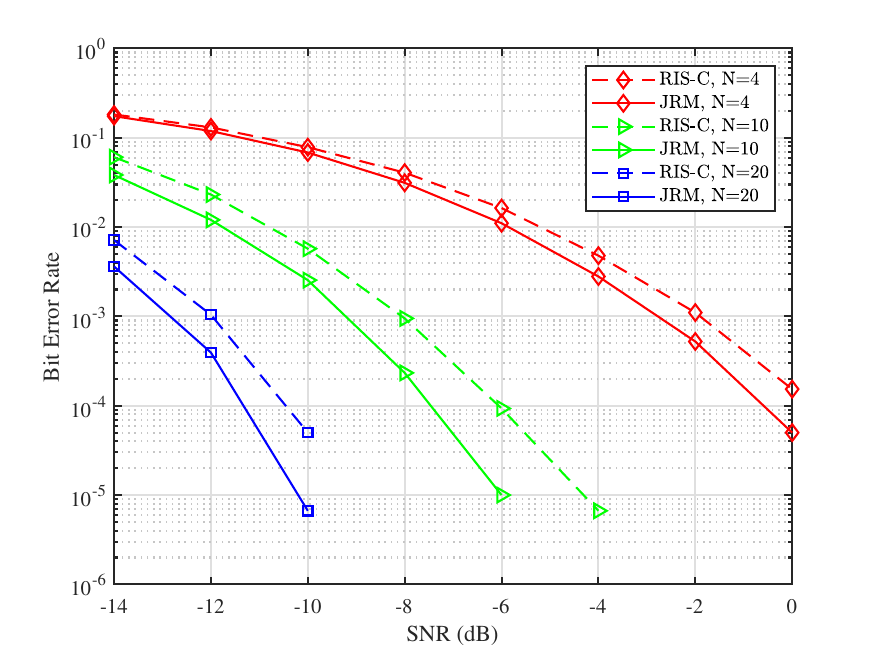}\\
 \caption{BER performance of the proposed JRM with the increase of the number of the RIS units with $N_t=2$, $N_r=3$ and $r=3$.}
  \label{result3c}
\end{figure}

Thirdly, we investigate the performance of the proposed schemes with the increase of the number of RIS units in systems  with $N_t=2$, $N_r=3$ and $r=3$, as illustrated in Fig. \ref{result3c}. With the increase of the number of RIS units from $4$ to $10$ and $20$, it is found that the system reliability is increased substantially by around $6$ dB and $11$ dB, respectively. This is because increasing the number of RIS units increases the number of reflected paths to the receiver. For comparison, the performance of RIS-C with the optimized fixed reflecting pattern is included. Simulation results demonstrate that JRM outperforms RIS-C under all system setups, which indicates that employing RIS as additional information modulators has performance advantages.

\subsection{Performance in Presence of Channel Estimation Errors}
\begin{figure}[t]
  \centering
  \includegraphics[width=0.5\textwidth]{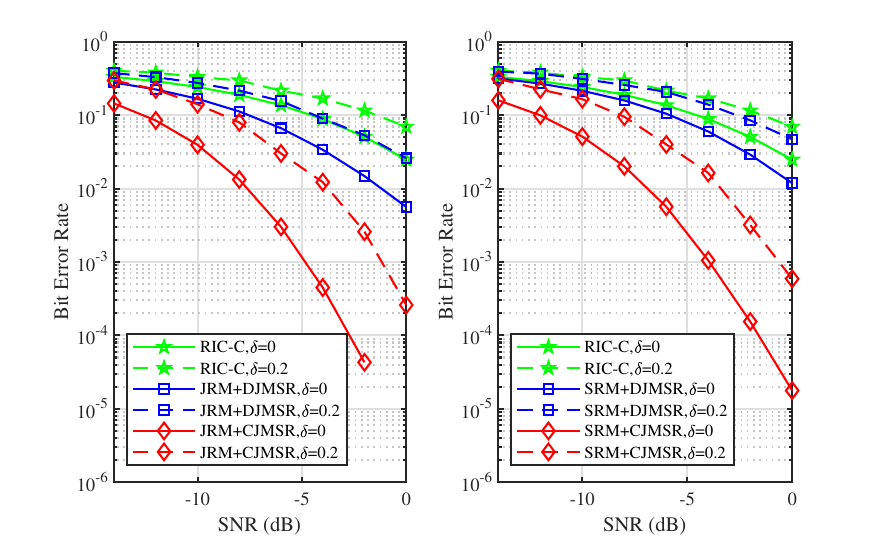}\\
 \caption{BER performance in the presence of channel estimation errors in a $(3,3,6,4)$ RIS-based MIMO communication system.}
  \label{result3}
\end{figure}

In the paper, a key assumption is that all CSI are globally and perfectly known by the transceivers and the RIS.  However, obtaining CSI is costly and the perfect CSI is typically not available. To investigate the applicability of the proposed designs to the scenarios with imperfect CSI. An channel estimation error model is adopted to model imperfect $\textbf{H}_d$, $\textbf{H}_1$ and $\textbf{H}_2$, which is written by $\textbf{H}_{\textrm{im}}=\textbf{H}_{\textrm{p}}+\textbf{H}_{\textrm{e}}$. In the model, $\textbf{H}_{\textrm{im}}$ and $\textbf{H}_{\textrm{p}}$ respectively denote the estimated and the real channel matrix; $\textbf{H}_e$ represents the channel error matrix with each entry following a zero-mean Gaussian variable with variance $(\delta\sigma)^2$ \cite{Guo2016,Guo2017,Guo2017a,Guo2019b}. This means that the channel estimation errors is propositional to the channel noise variance. Based on the channel estimation error model, we simulate the BER of the proposed designs under severe channel estimation errors and compared it with the design employing the perfect CSI. Simulation results as illustrated in Fig. \ref{result3} show that those designs are all affected by channel estimation errors.  Even though the proposed JRM\&SRM with CJMSR are more  sensitive to the channel estimation errors, they still show great performance improvement compared to RIS-C in the presence of the same level of channel estimation errors.

\section{Implementation Challenges and Future Directions}

The  implementation challenges of RM lies in that the synchronization between the transmitter and RIS, the fast reflecting modification at the RIS, the acquisition and sharing of precise CSI for DJMSR\&CJMSR. First, to encode the information together, the transmitter and RIS should be perfectly synchronized. This is challenging since RIS do not have information processing capacity on the surface. It requires additional synchronization signal receive circuit at the controller.  Second,  JRM\&SRM have to modify the reflecting patterns in a system time. It is challenging for high-data-rate short-symbol-time communication, since the fast modification may be limited by the hardware. But, it is worth mentioning that RIS-BC also needs to modify the reflecting patterns per symbol time and the data rates reported in current experimental work \cite{Basar2019a,Tang2019a, Tang2019b, Tang2019c} are growing higher. We believe that the problem can be addressed with the development of hardware. Third and most importantly, CSI acquisition and sharing are problematic. The design in this paper assuming each channels are perfectly known. However, due to the lack of signal processing capacity, the separate channels cannot be known. Thus, how to design the JRM\&SRM systems based on cascaded channel estimation and CSI sharing is a promising direction.  Besides, the information carrying capacity of JRM\&SRM should be probed in to provide design guideline and insights. JRM\&SRM-based multi-user communications call for a through and deep investigation.

\section{Conclusions}

This paper aimed to contribute to this growing area of research by exploring RIS for information transfer. A general concept named RM with more flexible transmit signals, reflecting patterns, and bit mapping methods was proposed. RM was classified into JRM and SRM (JRM\&SRM) according to the fact that the transmitter and RIS jointly transmit infomation or not. Both analysis and simulations results showed that the optimized JRM\&SRM surely outperform existing designs. To enhance transmission reliability, this paper proposed a discrete optimization-based joint signal mapping, shaping, and reflecting design for  JRM\&SRM to minimize the system BER with  a given transmit signal candidate set and a given reflecting pattern candidate set. To further improve the performance, we proposed to alternatively optimize the signal shaping and reflecting in continuous fields. In the reflecting design with given transmit signal sets, multiple reflecting patterns for reflecting and carrying information were jointly optimized.  In the signal shaping design with given reflecting patterns, the transmit signal sets for all reflecting patterns were jointly optimized. The computational complexity of all proposed designs were quantitatively in the number of multiplications. The performance of the proposed designs were investigated in various systems with and without considering CSI estimation errors. It was found that JRM\&SRM achieve much better performance than existing schemes in all cases. Moreover, implementation challenges and future directions of RM were pointed out. We believe that our proposed generalized transmission model and design methods can provide a significant reference to related research.

\bibliographystyle{IEEEtran}
\bibliography{IEEEabrv,RM}

\end{document}